\documentclass[manuscript]{aastex}

\def\kms{~km~s$^{-1}$\ }

\slugcomment{revised Dec, 08, 2011}

\shorttitle{Young Clusters in M31}
\shortauthors{Kang et al.}

\begin{document}

\title{A Comprehensive $GALEX$ Ultraviolet Catalog of Star Clusters in M31 and a Study of the Young Clusters}

\author{Yongbeom Kang\altaffilmark{1,2}, Soo-Chang Rey\altaffilmark{1,4}, Luciana Bianchi\altaffilmark{2}, Kyungsook Lee\altaffilmark{1}, YoungKwang Kim\altaffilmark{1}, and Sangmo Tony Sohn\altaffilmark{3}}
\altaffiltext{1}{Department of Astronomy and Space Science, Chungnam National University, Daejeon 305-764, Korea}
\email{ybkang@cnu.ac.kr, screy@cnu.ac.kr}
\altaffiltext{2}{Department of Physics and Astronomy, Johns Hopkins University, Baltimore, MD 21218, USA}
\altaffiltext{3}{Space Telescope Science Institute, Baltimore, MD 21218, USA}
\altaffiltext{4}{Author to whom any correspondence should be addressed}
 
\begin{abstract}

We present a comprehensive catalog of 700 confirmed star clusters in the field of M31 compiled from three major existing catalogs. 
We detect 418 and 257 star clusters in \textit{Galaxy Evolution Explorer} ($GALEX$) near-ultraviolet (NUV) and 
far-ultraviolet (FUV) imaging, respectively. 
Our final catalog includes photometry of star clusters in up to 16 passbands ranging from FUV to NIR 
as well as ancillary information such as reddening, metallicity, and radial velocities. 
In particular, this is the most extensive and updated catalog of UV integrated photometry for M31 star clusters. 
Ages and masses of star clusters are derived by fitting the multi-band photometry with model spectral energy 
distribution (SED); UV photometry enables more accurate age estimation of young clusters. 
Our catalog includes 182 young clusters with ages less than 1~Gyr. 
Our estimated ages and masses of young clusters are in good agreement with previously determined values in the literature. 
The mean age and mass of young clusters are about 300~Myr and 10$^{4}~M_{\sun}$, respectively. 
We found that the compiled [Fe/H] values of young clusters included in our catalog are systematically lower (by more than 1~dex) 
than those from recent high-quality spectroscopic data and our SED fitting result. 
We confirm that most of the young clusters kinematics show systematic rotation around the minor axis and 
association with the thin disk of M31. 
The young clusters distribution exhibits a distinct peak in the M31 disk around 10 - 12~kpc from the center and follow a spatial distributions 
similar to other tracers of disk structure such as OB stars, UV star-forming regions, and dust. 
Some young clusters also show concentration around the ring splitting regions found in the southern part of the M31 disk 
and most of them have systematically younger ($<$ 100~Myr) ages. 
Considering the kinematical properties and spatial distribution of young clusters, they might be associated with 
the well-known 10~kpc star-formation ring structure in the M31 disk. 
Consequently, we suggest that various properties of young clusters in M31 might be in line with the scenarios that 
a satellite galaxy had passed through the disk of M31 less than few hundred million years ago. 

\end{abstract}


\section{Introduction}

In the galaxy formation scenario based on the cold dark matter models, galaxies build up hierarchically
\citep{whi78,whi91,spr05}.
In this scenario, merging and accretion play key roles over cosmic time.
From high redshift to the nearby universe, many massive galaxies show evidence of ongoing merging and/or accretion.
In this context, large disk galaxies like the Milky Way (MW) and M31 are also thought to have assembled a significant
fraction of their mass through interactions with other small galaxies \citep[e.g.,][]{iba01}.
Most of these interactions change the morphology and star formation history of the galaxy.

As a typical spiral galaxy in the nearby Universe \citep{ham07}, M31 provides a unique and most
important opportunity for testing this scenario on external spiral galaxies due to its proximity \citep[785~kpc;][]{mcc05}.
Many recent studies have suggested that M31 is a promising example, exhibiting a hint of a past merger
\citep{blo06,brow06,gor06,iba07,mor08,mcc09}.
Most observational and theoretical results concern the halo and outer disk of M31. 
Especially, previous photometric and spectroscopic observations suggest that merging events have played an important
role in the construction and evolution of the halo of M31 (\citealt{mcc09} and references therein).

Recently, in addition to the well-known 10~kpc ring seen in previous observations 
(\citealt{gor06} and references therein), the presence of a second, inner dust ring was discovered in the disk
of M31 \citep{blo06}.  
While a detailed study of the origin of the ring structure of M31 is needed, the two off-center circular rings 
suggest that the disk of M31 has been distorted by a very recent passage of its satellite galaxy through the disk 
\citep[i.e., a head-on collision with the satellite galaxy about a few tens or hundreds Myr ago,][]{gor06,blo06}.
In this case, such a recent event may have enhanced the efficiency of star formation in the disk of M31 \citep[e.g.,][]{yin09}.

Star cluster systems can be a tracer of galaxy formation and assembly, in the sense that
significant star cluster formation is typically produced by major star-forming episodes in a galaxy
\citep{lar00,bro06}. 
More than 400 globular clusters (GCs) are known in M31 \citep[e.g.,][]{pea10}, which is about a factor of three more than in the MW. 
The GC system of M31 has two subpopulations, one is a metal-rich and spatially concentrated
subpopulation and the other is metal-poor and spatially extended. 
The metal-rich GCs show ``bulge-like" kinematics with rotation \citep{per02,lee08}. 
However, unlike in the MW, the metal-poor GCs also show significant rotation \citep{huc91,per02,lee08}.
\cite{mor04} showed a thin disk population of GCs which constitutes about 27~\% of the \cite{per02} sample.
Subsequently, it has been shown that at least a subset of these objects are in fact young ($\le$ 1~Gyr), 
metal-rich star clusters rather than old metal-poor GCs \citep{bea04,bur04,fus05,puz05,rey07,cal09,per10}.

In contrast to the MW, a large population of young clusters with ages less than 1 - 2~Gyr is found in M31
\citep{bur04,bea04,bea05,fus05,puz05,rey07,cal09,pea10}. 
\cite{fus05} presented 67 young clusters from the Revised Bologna Catalog \citep[RBC,][]{gal04} 
showing blue optical colors [$(B-V)_{0} < 0.45$] and/or high strength of H$\beta$ spectral index (H$\beta > 3.5$~\AA).
\cite{rey07} confirmed these young clusters using \textit{Galaxy Evolution Explorer} ($GALEX$) ultraviolet (UV) photometry and suggested that
the existence of young clusters in the outskirts of the M31 disk is due to the occurrence of significant
recent star formation in the thin-disk. 
More recently, two comprehensive catalogs of young clusters in M31 have been published from the spectroscopic 
survey of \cite{cal09} and the Sloan Digital Sky Survey (SDSS) data of \cite{pea10}. 
\cite{cal09} estimated ages and masses of 140 young clusters and \cite{pea10} defined 
156 young clusters with blue colors of $g-r < 0.3$. 
Most of these clusters are more massive (between 10$^3$ and 10$^5$~$M_\sun$) than the Galactic open clusters \citep{cal09}.
Furthermore, they have similar characteristics to the blue star clusters in the LMC \citep{bur04,fus05} 
and other massive young clusters in Local Group galaxies \citep{bar09}. 
However, no such predominant counterparts have yet been discovered in the disk of the MW, except for a handful of massive young clusters identified 
in the center of the MW \citep[e.g.,][]{mes09}. 
The existence of massive young clusters in the outskirts of the M31 disk indicates the occurrence of significant
recent star formation in the disk of M31 \citep{fus05,rey07,cal09}. 
Assuming that merging and accretion event triggered higher-level star formation in the disk of M31 than in quiescent
galactic disks, it is interesting to examine the properties of star clusters related to the M31 disk,
elucidating the recent star formation history in M31.

Motivated by the opportunity to study formation and evolution of young clusters in M31, in this paper we construct 
a comprehensive multi-band catalog of star clusters in M31 compiled from RBC, \cite{cal09,cal11}, and \cite{pea10} samples. 
In particular, we included $GALEX$ UV data, since the UV flux is highly sensitive to young main-sequence stars included 
in the massive young clusters which radiate more UV flux than in optical passbands \citep{rey07,kav07}. 
We select extensive subsamples of young clusters which is complementary with previous catalogs of young clusters 
in M31 \citep[e.g.,][]{cal09}. 
Various properties (age, mass, metallicity, kinematics, and spatial distribution) of young clusters are compared with 
star-forming (SF) regions and OB type stars in M31, and with the 10 kpc ring structure. 
This allows us to test whether most young clusters may be the possible outgrowth of a recent accretion of 
satellite galaxy occurred at the center of the M31 disk.

In Section 2, we describe optical and near-infrared (NIR) data of the M31 star clusters compiled from previous catalogs. 
Combining additional $GALEX$ UV data and other auxiliary information, we present a final merged catalog of star clusters in M31. 
In Section 3, by comparing observed spectral energy distributions (SEDs) with simple stellar population models, 
we estimate ages and masses of star clusters and select young clusters. 
We present properties of young clusters in Section 4. 
We also discuss the relationship between young clusters and M31 disk structures in terms of possible recent star formation 
history in the M31 disk. 
The conclusions are summarized in Section 5.

\section{Data}

\subsection{Photometric Data}

\subsubsection{Optical and Near-infrared Data}

Large catalogs of star clusters in M31 have been published in the past decade \citep[e.g.,][]{bar00,gal04,kim07,cal09,cal11,pea10}.
However, it is still a challenge to build a complete, deep, and homogeneous catalog of star clusters in M31. 
For example, it is not easy to detect relatively faint star clusters which are mainly projected onto the bright disk 
structure or bulge of M31. 
Furthermore, some of the star clusters exist in the halo, far away from the host galaxy \citep{hux08}, 
requiring wide-field surveys of the outer halo of M31.

One of the most self-consistent catalogs of star clusters in M31 is that of \cite{bar00}, in which they 
presented $UBVRI$ and $JHK$ photometry of 435 clusters and cluster candidates. 
However, only for 268 objects optical photometry in four or more bands is available, and 224 have NIR photometry. 
\cite{gal04} identified 693 clusters and cluster candidates from the 2 Micron All Sky Survey (2MASS) database. 
They provide an extensive RBC which includes compiled multi-band optical data from many previous catalogs. 
\cite{kim07} carried out wide field observations and found 113 new genuine star clusters and 258 probable star clusters 
which are mostly faint (18 $\le V \le$ 20~mag) objects. 
\cite{cal09} published a new catalog of 670 likely star clusters, with accurate coordinates from the 
Local Group Galaxy Survey \citep[LGGS,][]{mas06} and the Digitized Sky Survey (DSS) data. 
Most of these clusters are confirmed based on high-quality spectra taken with the Hectospec spectrograph on the 6.5~m MMT. 
These authors also estimated ages, reddening values, and masses of 140 young clusters by comparing the observed spectra with model ones. 
They presented only $V$-band photometry for 510 clusters from the LGGS images, with no information on colors. 
Based on the classification of \cite{cal09}, subsequently \cite{cal11} also provided metallicities and ages of 
367 old clusters using the high-quality spectra. 
\cite{pea10} presented an updated catalog including new, consistent $ugriz$ and $K$-band photometry 
based on images from the SDSS and Wide Field CAMera (WFCAM) on the UK Infrared Telescope (UKIRT). 
This catalog includes homogeneous photometry of 572 clusters and 373 candidate clusters. 
Using archival images from the LGGS, \cite{fan10} recently presented an updated $UBVRI$ photometric catalog containing 
970 objects selected from the RBC. 

For our following analysis, we construct a new compiled catalog of star clusters in M31 carefully considering 
three previously published catalogs: RBC version 4 (v4), \cite{cal09,cal11}, and \cite{pea10}. 
Our catalog is mainly based on the RBC v4 which is the most extensive and commonly used catalog although it contains 
rather heterogeneous photometry compiled from various literature \citep{gal04}. 
As of December 2009, the RBC v4 includes most previous data, except for the catalogs of 
\cite{pea10} and \cite{cal11}, and contains 667 star clusters and 606 candidate clusters. 
We carefully compared names and coordinates of star clusters between the RBC v4, \cite{cal09,cal11}, and \cite{pea10} catalogs. 
While most objects are well matched in these catalogs, some have slightly different coordinates in the RBC v4. 
From inspection of LGGS and SDSS images, we found the coordinates of 17 objects provided by RBC v4 to be 
discrepant with real centers of the objects and finalized their coordinates with those of \cite{cal09,cal11}. 
These are  B284, B353, B414, NB18, NB42, NB44, NB104, B001D, B003D, B246D, B306D, DAO89, V203, M075, M088, BH01, BH07. 
We add 56 objects from the catalogs of \cite{cal09,cal11} and \cite{pea10} which are not contained in the RBC v4. 
These are previously known objects however they are not included in the RBC v4.

The final compiled catalog contains a total of 2,101 objects. 
This catalog contains star clusters, candidate clusters, HII regions, stars, asterisms, and background galaxies 
classified from RBC v4, \cite{cal09,cal11}, and \cite{pea10}. 
Some objects do not have the same classification in these three catalogs. 
In this paper, we only consider 700 star clusters which are classified as confirmed star clusters at least in one of three catalogs. 
As a result our compiled catalog is the most extensive one for confirmed star clusters in M31 (see Section 2.3 and Table 1).

\subsubsection{$GALEX$ Ultraviolet Data}

UV data is very powerful tool for breaking age-metallicity degeneracy and estimating the ages of star clusters \citep{kav07,bia09,bia11}. 
We used UV images from the Nearby Galaxy Survey (NGS) obtained by $GALEX$ in two UV bands: far-ultraviolet 
(FUV; 1350 - 1750~\AA) and near-ultraviolet (NUV; 1750 - 2750~\AA). 
Every $GALEX$ image has 1.25~deg circular field of view \citep{mor07}. 
A total of 23 images (about 17 square degrees) have covered most of the disk and halo of M31. 
The details of the $GALEX$ observations for M31 are presented in \cite{rey05,rey07}.

Of each image, we only use the inner 1.1~deg field to avoid the distortion at the edge of the field. 
Aperture photometry of all point sources in the M31 fields was carried out using the DAOPHOTII package \citep{ste87}. 
We measured the flux of each source within 3~pixel (4.5~arcsec) radius and applied aperture correction using isolated stars in each image. 
The measured fluxes were converted to the AB magnitude system with the calibration of \cite{mor07}. 
Our UV photometry is the same as that published by \cite{rey07}.
Sources in our $GALEX$ photometry were cross-matched with clusters in our compiled catalog using a matching radius of 6~arcsec. 
We then carried out careful visual inspection of all matched objects in each $GALEX$ image and reject all spurious sources  
(i.e., sources highly contaminated by nearby objects, faint fuzzy sources, and noisy pixels). 
Out of the 700 star clusters in the compiled catalog, 418 ($\sim$60~\%) and 257 ($\sim$37~\%) objects are detected in the 
$GALEX$ NUV and FUV band, respectively. 
Of these, 302 and 167 objects were detected in the previous NUV and FUV data of \cite{rey07}. 
The limiting magnitudes of star clusters are 23.6~mag and 23.7~mag for FUV and NUV, respectively. 

In Figure 1, we present the spatial distribution of the star clusters detected in $GALEX$ NUV and FUV bands 
with respect to the M31 disk, NGC 205, and M32. 
We examine the detection rate of star clusters in our $GALEX$ fields with respect to their $B$ and $V$ magnitudes. 
Figure 2 shows the fraction of star clusters detected in the NUV and FUV bands as a function of $V$ magnitude and $B-V$ color. 
Of the 484 clusters with both $B$ and $V$ data, 328 (about 68~\%) and 191 (about 39~\%) clusters are detected in the 
$GALEX$ NUV and FUV, respectively. 
The color-magnitude diagrams and color histogram show that most of the detected objects are optically blue clusters with $B-V < 1.2$. 
Many of the bluest clusters with $B-V < 0.5$ are detected in the $GALEX$ UV bands even though 
they are fainter ($V > 16$) than the redder clusters in the optical passband. 
Most of these blue clusters are young clusters (see Section 3).

\subsection{Auxiliary Data: Reddening, Metallicity, and Radial Velocity}

We used reddening values of star clusters from \cite{bar00}, \cite{fan08}, and \cite{cal09,cal11}. 
\cite{bar00} and \cite{fan08} estimated the reddening values from reddening-free parameters 
and color-metallicity relation. 
\cite{cal09,cal11} published reddening values of young and old clusters, separately, 
which were derived by comparing the observed spectra with model ones. 
The mean differences between reddening values of \cite{cal09,cal11} and those of \cite{bar00} and \cite{fan08} are +0.03 for both cases. 
In the case of star clusters with available reddening values in more than two different works, the average value 
has been adopted. 
The reddening values are available for 555 star clusters in our compiled catalog.

As for the 145 star clusters with no available reddening values in the literature, using 555 star clusters with 
available reddening values, we calculate median reddening values of star clusters located within an annulus at every 2~kpc 
radius from the center of M31. 
Figure 3 shows the distribution of our compiled reddening values of 555 star clusters and calculated median reddening values of each annulus. 
As we can expect, the reddening values are maximum around the 10~kpc star formation ring in the M31 disk. 
Finally, we adopt the median reddening value of each annulus for star clusters with no available reddening estimates. 
However, beyond a galactocentric distance of 22~kpc, the reddening values converge to $E(B-V)$ = 0.13~mag which is similar to the mode value of 
all old GCs of \cite{cal11}. 
We adopt the $E(B-V) = 0.13$~mag for the star clusters at distances larger than 22~kpc, since most of them are located in halo regions.

From their Hectospec spectroscopic observations, \cite{cal11} presented metallicity values of 333 old clusters. 
Currently, this is the most extensive and homogeneous metallicity data-set. 
We adopt their metallicity values as a basic data, and also combine other metallicity values from \cite{gal09}, \cite{per02}, and \cite{bar00}. 
The mean differences between metallicity values of \cite{cal11} and others are $-$0.07 for \cite{gal09}, +0.05 
for \cite{per02}, and +0.14 for \cite{bar00}. 
Finally, we adopt the mean value of metallicity from these works. 
For star clusters with metallicity value in only one paper, we adopt that value. 
The metallicity values are available for 399 star clusters in our compiled catalog.

The RBC v4 includes radial velocities of 528 star clusters as weighted mean values from various literature 
\citep[e.g.,][]{bar00,per02,gal06,kim07,cal09,alv09}.
However, RBC v4 was not updated with the most recent data by \cite{cal11} for 507 star clusters. 
The mean difference between radial velocities of the RBC v4 and \cite{cal11} is about 3~\kms. 
Finally, we adopt the mean value of radial velocity from these two catalogs when more than one measurement is available. 
The final radial velocities are available for 617 star clusters in our compiled catalog.

\subsection{Merged Catalog of Star Clusters in M31}

Our final merged catalog of M31 star clusters is presented in Table 1. 
This catalog includes 700 star clusters with photometry in up to 16 passbands ranging from FUV to NIR as well as 
ancillary information such as reddening values, metallicity values, and radial velocities. 
Optical $U, B, V, R, I$ and NIR $J, H, K$ bands are from the RBC v4. Optical $u, g, r, i, z$ and NIR $K$ bands are from \cite{pea10}. 
Our compiled catalog is then used for the selection and analysis of young clusters in the following Sections. 
This is the most extensive and updated catalog of UV photometry for M31 star clusters, superseding 
our previous UV catalog \citep{rey07}. 
The following is a brief description of each column of Table 1;

\begin{itemize}

\item $Column$ (1): name of star cluster

\item $Column$ (2) - (3): coordinates of star cluster (hh:mm:ss, dd:mm:ss) 

\item $Column$ (4): $GALEX$ FUV magnitude (AB mag)

\item $Column$ (5): uncertainty of FUV magnitude (AB mag)

\item $Column$ (6): $GALEX$ NUV magnitude (AB mag)

\item $Column$ (7): uncertainty of NUV magnitude (AB mag)

\item $Column$ (8) - (15): $U, B, V, R, I, J, H, K$ magnitudes from RBC v4 (VEGA mag)

\item $Column$ (16) - (20): $u, g, r, i, z$ magnitudes from \cite{pea10} (AB mag)

\item $Column$ (21): $K$ magnitude from \cite{pea10} (VEGA mag)

\item $Column$ (22): reddening value 

\item $Column$ (23): uncertainty of reddening value, obtained from compilation of different sources 

\item $Column$ (24): [Fe/H] value

\item $Column$ (25): uncertainty of [Fe/H] value

\item $Column$ (26): radial velocity (\kms)

\item $Column$ (27): uncertainty of radial velocity (\kms)

\item $Column$ (28): classification flag of RBC v4 (1: cluster, 2: candidate cluster, 3: controversial object, 4: galaxy, 5: HII region, 6: star, 7: asterism, 8:extended cluster, 99: no data)

\item $Column$ (29): classification flag of \cite{pea10} (1: old cluster, 2: candidate cluster, 3: young cluster, 4: galaxy, 5: HII region, 6: star, 99: no data)
 
\item $Column$ (30): classification flag of \cite{cal09,cal11} (1: young cluster (age $<$ 1~Gyr), 2: intermediate cluster (1 $<$ age $<$ 2~Gyr), 3: old cluster (age $>$ 2~Gyr), 4: cluster (no age), 5: star, 6: maybe star, 7: HII region, 8: unknown, 9: candidate cluster, 10: weird (SNR, Eta Carina type, or symbiotic star), 99: no data)

\item $Column$ (31): flag of $E(B-V)$ (1: mean value of reddening from available literature, 2: median reddening value of star clusters located within an annulus at every 2~kpc radius from the center of M31,  3: foreground reddening value of $E(B-V)=0.13$~mag)
\end{itemize}

\section{Selection of Young Clusters}

\subsection{Multi-band SED Fitting of Star Clusters}

In order to estimate accurate ages and masses of the star clusters using multi-band SED fitting, 
we need many photometric data points covering as wide a wavelength range as possible 
\citep{deg03,and04,kav07,wan10,fan10}.
In particular, UV photometry with optical one produces age constraints comparable to those of spectroscopic observations \citep{kav07}. 
Our compiled catalog includes extensive photometric data in 16 bands from FUV to NIR. 
On the other hand, the photometric uncertainties are also important. 
Since the $UBVRI$ data are from RBC v4, we adopt photometric uncertainties following \cite{gal04}, 
i.e., $\pm$0.05~mag in $BVRI$ and $\pm$0.08~mag in $U$. 
We also adopt a typical error ($\pm$0.05~mag) for the $JHK$ data from RBC v4. 
\cite{pea10} present $ugriz$ and $K$ data and their photometric uncertainties. 
However, most of these uncertainties are extremely small, therefore we added 
a 0.05~mag uncertainty in all bands to account for zero-point inconsistencies among 
the various datasets. 
The photometric uncertainties of our $GALEX$ FUV and NUV data are included in Table 1.

Basically, the multi-band SED fitting method is a comparison between multi-band photometry and synthetic model magnitudes 
of simple stellar population (SSP). 
A SSP is defined as a single generation of coeval stars characterized by fixed parameters such as 
metallicity, age, and stellar initial mass function (IMF). 
Synthetic SSP models are calculated on the basis of a set of evolutionary tracks of stars of different 
initial masses, combined with stellar spectra at different evolutionary stages. 
In this paper, we compare the multi-band SEDs of our star clusters with magnitudes constructed from progressively reddened 
PADUA SSP models \citep[see][]{bia11} to estimate their ages. 
We use a \cite{sal55} IMF with lower and upper mass limits of 0.10~$M_{\sun}$ and 100~$M_{\sun}$. 
After age and extinction are constrained from the SED colors, scaling the best-fit model to the observed magnitudes yields 
the cluster mass, since the distance is known. 

Since the reddening is critical for an accurate age estimation of star clusters, 
we explored two ways for adopting a final reddening value. 
First, we adopted the reddening value from our merged catalog (``$indivEBV$") and only derived the cluster age from SED fitting, 
imposing the adopted $E(B-V)$. 
The second way was to treat both age and $E(B-V)$ as free parameters in the SED fitting analysis (``$freeEBV$"). 
The $freeEBV$ is important for the 145 star clusters whose reddening values are not available in the 
literature (see Section 2.2). 
Since the UV extinction curve of M31 is similar to that of MW \citep{bia96}, we assumed MW-type interstellar 
reddening ($R_V$ = 3.1) (see \citealt{kan09} for a discussion of the effects of different extinction curves). 
For our SED fitting analysis, we used model grids with five different metallicities, Z = 0.0004, 0.004, 0.008, 0.02 (solar metallicity), 
and 0.05, although M31 is believed to have a higher metallicity than the MW (e.g., \citealt{mas03} and references therein). 
In this paper, we did not consider metallicities lower than Z = 0.0004, since we focus on the young clusters which are mostly 
metal-rich with [Fe/H] $>$ $-1.0$ (see Section 4.2). 
We ran the SED fittings using each metallicity for all star clusters, in order to assess the effects of this parameter. 

We compared the $freeEBV$ results from SED-fitting with the literature values of $E(B-V)$ ($indivEBV$). 
There is good agreement for part of the sample (28~\% of the whole final sample have $E(B-V)$ in agreement within $\Delta E(B-V) = 0.1$~mag) 
while other sources have larger discrepancies. 
The $freeEBV$ values tend to be lower than literature values, but the mean difference is not significant 
(mean $\Delta E(B-V)$ $\sim$0.09~mag with a $\sigma$ of 0.19~mag). 
The adopted extinction curve, and model details, may also affect the results. 
More important for our purpose is the effect of the uncertainty in $E(B-V)$ on the derived ages.
We must recall first of all that for some of the catalog magnitudes compiled and used here, no errors are reported, 
which prevents derivation of formal errors from the SED fitting procedure (a constant uncertainty of some reasonable value 
has to be assumed for the $UBVRIJHK$ photometry). 
Therefore, we estimated the robustness of the SED-based results by comparison with $E(B-V)$ from previous works. 
Figure 6 shows how the $E(B-V)$ uncertainty affects the derived ages. 
Where there is agreement in the derived $E(B-V)$ (e.g., within 0.1~mag, for 37~\% of the clusters if we used the results from FN$ugriz$ assuming Z=0.02), obviously ages are in good agreement. 
In some cases where high discrepancies are seen, these may be due to some mismatch in SEDs between instruments 
(for example due to nearby objects), or to several minima being possible in the SED fitting.
Overall, we see that the extraction of the "young clusters" subsample is not affected by these uncertainties.

In our SED fitting, we separated $UBVRI$ data from $ugriz$ data in order to keep the homogeneity of optical data. 
We also separated $JHK$ (RBC v4) from $K$ \citep{pea10} data. 
We considered four combinations of photometric passbands for our SED fitting. 
We include our UV data in each case. 
They are: (1) FUV, NUV, and $ugriz$ (``FN$ugriz$"), (2) FUV, NUV, and $UBVRI$ (``FN$UBVRI$"), 
(3) FUV, NUV, $ugriz$, and $K$ (``FN$ugrizK$"), and (4) FUV, NUV, $UBVRI$, and $JHK$ (``FN$UBVRIJHK$"). 
In Figure 4, we present a sample (B100) of SED fitting results with four different band combinations. 
The estimated ages are similar between four different results. 
However, in many cases of FN$UBVRIJHK$, the photometry in $JHK$ bands shows obvious offsets from UV and optical bands, probably due to 
the low resolution of 2MASS (see Figure 5). 
$GALEX$ resolution is also low ($\sim 5~\arcsec$), UV-bright objects are rare, compared with IR sources, and contamination is less 
likely, although possible. 
The FN$ugrizK$ provides a homogeneous dataset, therefore we did not consider the FN$UBVRIJHK$ in our final SED fitting analysis. 
SED analysis requires photometric measurements in at least three passbands.
We run the SED fitting for combinations of five different metallicities, two different reddening treatments, and 
three photometric combinations of different passbands. 
Based on the $\chi^2$ minimization result, we computed the best age, or [age, $E(B-V)$] combination, and the uncertainty 
in the derived value from the $\chi^2$ contours equal to minimum $(\chi^2)+1$. 
We also computed the probability of the solution from each run to be the most appropriate one, given by 
a likelihood estimator of the form $p \sim exp(-\chi^2)$. 
Of the 30 different estimates of ages and masses of each star cluster from our SED fitting analysis, 
we select final values with highest fitting probability. 
The typical uncertainty in age and mass is about 33~\% across the whole sample, but smaller for younger clusters: 
16~\% for the subsample with ages $<$ 1~Gyr, and 44~\% for the older clusters.
These errors are the formal uncertainties from the SED-fitting with best metallicity and extinction chosen in each case. 
We recall however, that for some of the photometry no errors were available and a constant uncertainty had to be assumed 
for the whole catalog; in addition, the derived mass uncertainty takes into account the photometric errors 
and the derived [age, $E(B-V)$]: when the extinction is high, a larger uncertainty might affect the estimated mass. 
The scatter between age and mass values from the best SED-fitting solution (using the chosen metallicity and reddening) and 
values obtained  with  different assumptions, gives an indication of possible additional uncertainties. 
The difference between ages derived using models with solar, versus Z=0.05 metallicity, is $\leq$ 50~\% 
at about 1~Gyr, $\leq$ 30~\% at 100~Myr and much less for younger ages, higher metallicity yielding younger ages. 
The difference in resulting age using models with Z=0.08 versus solar metallicity is somewhat smaller, 
and between solutions with Z=0.008 versus Z=0.004 is much smaller. 
Precise estimates of metallicity from spectroscopy would be relevant to eliminate these factors of uncertainty.

We obtained results for 403, 185, and 57 objects from FN$ugriz$, FN$UBVRI$, and FN$ugrizK$, respectively. 
In most cases, homogeneous optical bands (e.g., $ugriz$ data) provided the best fits (see Figure 4). 
For 55 objects with measurements in less than three bands we do not estimate ages. 
The reddening values of 409 objects are adopted from the $indivEBV$ and those of 236 objects are from the $freeEBV$. 
In Figure 7, we present the distribution of our estimated  ages and masses of star clusters in M31. 
Comparison with young ($\le$ 1~Gyr; blue filled circles) and old ($>$ 1~Gyr; red filled circles) star clusters  
obtained from \cite{cal09,cal11} shows that our estimations are  consistent with their results.  
Finally, we obtained ages and masses of 176 young ($\le$ 1~Gyr) clusters and 446 old ($>$ 1~Gyr) clusters from our analysis.

\subsection{Comparison with Previous Results}

In Figure 8, we compare our estimated ages of young clusters with results from previous works 
\citep[e.g.,][]{bea04,puz05,van09,wan10,fan10,per10,cal09,cal11}. 
\cite{bea04} estimated ages of 8 young clusters from a comparison of observed spectra with synthetic SSP models. 
Their ages are in good agreement with our estimations although there is a small ($\sim$0.1~Gyr) systematic offset. 
\cite{van09} estimated ages of star clusters located in the southern disk of M31 by multi-band ($UBVRI$) SED fitting. 
Our estimated ages of young clusters are similar to their results, albeit with some scatters. 
\cite{puz05} compared the Lick indices to SSP models for their age estimation. 
Their ages show large discrepancy from ours.  
Since the SSP models which they used do not cover the age range less than 1~Gyr, the model limitation might be responsible 
for this discrepancy in young clusters. 
\cite{per10} estimated ages of young clusters by comparing model isochrones with color-magnitude diagrams obtained 
from HST/WFPC2 observations. 
Their results are also in good agreement with ours. 
\cite{wan10} and \cite{fan10} estimated ages of clusters from multi-band SED fitting. 
Many of their estimated ages are largely discrepant from ours. 
Most of young clusters identified by them are in fact old and metal-poor \citep[see also][]{cal11}.
\cite{cal09,cal11} published ages of star clusters by comparing their high-quality integrated spectra with SSP models. 
Our age estimation for young clusters are in good agreement with results of \cite{cal09,cal11}. 
It is worth to note that our age estimation based on SED fitting of multi-band photometry including UV data is comparable 
to those achieved by other works using spectroscopic line indices and color-magnitude diagrams. 
This emphasizes again that the UV photometry is a powerful tool for age estimation of young stellar populations \citep[see also][]{kav07,bia11}.

In Figure 9, we also compare masses of young clusters from our analysis with other works: 
\cite{bea04}, \cite{van09}, \cite{per10}, and \cite{cal09,cal11}.
In previous studies, masses are estimated from the mass-to-light ratio ($M/L$) coupled with estimated ages. 
Our estimated masses are slightly larger than those of \cite{bea04}, \cite{van09}, and \cite{per10} 
by factors of 1.7, 1.8, and 1.6, respectively. 
However, our masses are in good agreement with estimations from \cite{cal09,cal11} (about 30~\% higher, on average). 
Our young clusters have masses in the range of $\sim3\times10^{2}$ - $2\times10^{5}$~$M_{\sun}$.

\subsection{Young Cluster Catalog}

From our SED fitting analysis in Section 3.1, we select 176 young clusters younger than 1~Gyr, 
and confirm that their ages and masses are in good agreement with other previous results (see Section 3.2). 
For a complete list of young clusters, in addition to our sample, we also consider the 155 young clusters 
with $\le$ 1~Gyr from \cite{cal09}. 
Among our 176 young clusters, we only include 173; 129 objects are classified as young clusters in both 
of ours and \cite{cal09} analysis and for 44 clusters, ages were not estimated by \cite{cal09}. 
We exclude three young clusters (B100, M019, PHF7-1) which have old ($>$ 1~Gyr) ages in \cite{cal09}. 
We add 9 young clusters which are only available in \cite{cal09} and adopted their ages and masses from \cite{cal09}. 
We exclude 17 young clusters from \cite{cal09} which are estimated to be older than 1~Gyr by our analysis. 
Finally, we construct a final catalog of 182 young clusters consisting of our 173 young clusters and 9 young clusters 
from \cite{cal09}. 
Table 2 presents a catalog of these young clusters with their ages and masses. 

\subsection{Color-Color and Color-Magnitude Diagrams}

Figure 10 shows extinction-corrected $g-r$ vs. UV$-r$ (upper two panels) and NUV$-r$ vs. FUV$-r$ (lower right panel) 
diagrams for star clusters included in our catalog. Blue filled circles and red filled circles are young ($\le$ 1~Gyr) 
and old ($>$ 1~Gyr) clusters, respectively. 
The mean distribution of young clusters is biased towards blue colors in both optical and UV colors. 
The discrimination between young and old clusters is more prominent in the diagram (NUV$-r$)$_0$ vs. (FUV$-r$)$_0$, 
since UV fluxes are sensitive to the young stellar populations. 
The old GCs in the MW \citep[crosses; Table 6 of][]{soh06} and the old clusters in M31 lay on the same locus 
in these diagrams. 
Different solid curves show Yonsei evolutionary population models in the age range 0.1 - 14~Gyr 
(from lower to upper: C. Chung et al. in preparation). 
The Galactic and M31 clusters follow the general trend indicated by the model lines.
 
In the $g-r$ vs. UV$-r$ diagrams, the dashed horizontal line corresponds to the reference value of $(g-r)_0$ = 0.3 
for young cluster selection adopted by \cite{pea10}. 
The dashed vertical lines are also arbitrary reference values of (NUV$-r$)$_0$ = 2.5 and (FUV$-r$)$_0$ = 3.0 for 
young cluster selection \citep{boh93,rey07}. 
In the ranges of (NUV$-r$)$_0$ $<$  2.5 and (FUV$-r$)$_0$ $<$ 3.0, the MW GC system lacks young clusters. 
Most of our young clusters have smaller values than the reference colors; $(g-r)_0$ $<$ 0.3, (NUV$-r$)$_0$ $<$ 2.5, and (FUV$-r$)$_0$ $<$ 3.0. 
\cite{fus05} selected massive young clusters in M31 according to their color and/or the strength of 
H$\beta$ spectral index with H$\beta$ $>$ 3.5~\AA. 
In Figure 10, we also denote the clusters with H$\beta$ $>$ 3.5~\AA~(open squares) compiled from \cite{bea04}, \cite{fus05}, and \cite{gal09}. 
It is clear that the distribution of H$\beta$-selected sample is consistent with that of our young clusters. 
The lower left panel of Figure 10 shows the $M_r$ vs. NUV$-r$ diagram for M31 clusters and MW GCs. 
A distance modulus of $(m - M)_0$ = 24.47 \citep{mcc05} was adopted for all M31 clusters. 
The most distinct feature is that young clusters with $\le$ 1~Gyr are systematically fainter in $V$ than the old 
clusters, which indicates that the M31 young clusters are systematically less massive objects than the old GCs in 
the M31 and the MW (see Section 4).

\section{Properties of Young Clusters}

\subsection{Age and Mass Distribution}

Figure 11 shows the distribution of estimated ages and masses for 182 young clusters. 
One interesting feature is that the majority of M31 clusters with age $<$ 1~Gyr is rather biased towards the younger age range of $<$ 500~Myr. 
In the age histogram (upper histogram), as the age of clusters decreases, the fraction of young clusters increases. 
About 82~\% (149 of 182) of the clusters are younger than 500~Myr. 
Clusters older than 500~Myrs with mass lower than $\sim10^{4}~M_{\sun}$ are too faint to be detected in the catalogs compiled in this work. 
However, even when we consider only clusters more massive than $10^{4}~M_{\sun}$, 73~\% (73 out of 100) are younger than 500~Myrs.
This may reflect effects of luminosity fading with age, cluster disruption, and possible variations in time of the cluster formation rate. 
On the other hand, most young clusters are in the mass range of $10^{3.5}$ - $10^{4.5}~M_{\sun}$. 
The mean value of age and mass of young clusters is about 300~Myr and $10^{4}~M_{\sun}$, respectively. 
Note that, in the same age range, the mass distribution of our young clusters in M31 is similar to 
that of massive clusters found in the LMC and is in between those of Galactic young clusters and 
old GCs \citep{bea04,fus05,cal09}. 
In Figure 11, we also note that there is a general lack of massive young clusters with  $>$ $10^{5}~M_{\sun}$ 
in M31. 

Interestingly, many young clusters younger than 50~Myr are low mass ones with $<$ $10^{3.5}~M_{\sun}$. 
Their masses are comparable to the mass range of typical MW open clusters \citep[see Figure 14 of][]{cal09}. 
Even though there are a handful of young clusters with very low masses ($<$ $10^{3}~M_{\sun}$) and ages less than 10~Myr, 
it is obvious that our sample is not complete in detecting such faint and low mass clusters.  
\cite{kri07} estimate that the entire disk of M31 contains approximately 80,000 such faint and small clusters 
extrapolating from their detected 343 clusters. 
Further deep HST observations for an extensive area of the M31 disk will clarify the nature of these low mass 
clusters and mass distribution of cluster system in M31 (e.g., Panchromatic Hubble Andromeda Treasury (PHAT) survey, 
\citealt{joh11}; Dalcanton et al. in preparation).

Young clusters are experiencing a serious loss of gas and dust during the supernovae explosion
phase \citep[10 - 50~Myr:][]{lad03,fal05,goo09}, internal dynamical evolution and stellar population fading 
 \citep[10 - 100~Myr:][]{lam09}, and galactic tidal effects and other external effects \citep[100 - 1000~Myr:][]{bou03,gie05,lam05,lam06,gie07}.
Many low mass clusters can become gravitationally unbound and easy to disrupt within these phases \citep{pel10}. 
While survival of star clusters depends upon the mass, size, and environment, most low mass young clusters 
found in the M31 disk might be disrupted within few Gyrs.

\subsection{Metallicity}

Most previous results concerning the metallicity of star clusters in M31 were focused on old GCs. 
In our compiled catalog, metallicity values are available for 46 young clusters (see Section 2.2). 
As shown in Figure 12, young clusters appear to be biased towards the metal-poor range of [Fe/H] $<$ $-$1.0. 
Our compiled metallicity values of 30 young clusters are from \cite{per02}, eight values are the mean value 
from \cite{per02} and \cite{bar00}, four values are from \cite{bar00}, and the remaining four values are from \cite{gal09}. 
All of these metallicities are estimated from the Lick indices which were calibrated from the Galactic 
old GCs. 
On the other hand, \cite{fus05} claimed that young clusters are probably not so metal-poor as deduced 
from the metallicity values provided by \cite{per02}. 
From the specific comparison between [Fe/H] values derived from different Lick indices, 
\cite{fus05} concluded that $G$-band line strength tends to underestimate [Fe/H] values 
in \cite{per02} by more than 1~dex. 

As a further argument, we compared metallicities of young clusters from our compiled catalog with 
other independent results from \cite{bea04} and \cite{per10}. 
\cite{bea04} obtained high-quality spectra for 8 young clusters and estimated their metallicities. 
Using HST/WFPC2 data, \cite{per10} derived ages and metallicities of young cluster by fitting 
optical color-magnitude diagrams with theoretical isochrones. 
The [Fe/H] values from \cite{bea04} and \cite{per10} are systematically higher (about 1.3~dex) than our compiled values. 
While the [Fe/H] values of young clusters in our catalog mainly rely on the estimation of 
\cite{per02}, we suggest these values might be underestimated. 
For completeness, in Table 1 ([Fe/H] values in parenthesis of column (24)), we also include metallicity values 
of 34 young clusters from \cite{bea04} and \cite{per10}. 

In our SED fitting analysis, we used five discrete metallicity grids (Z = 0.0004, 0.004, 0.008, 0.02, 0.05) 
of SSP models. 
Although accurate metallicities cannot be derived from SED fitting, we can chose the model grid which provides the best fit, as an indication. 
In Figure 13, we compare our compiled [Fe/H] with results from our SED fitting. 
Our [Fe/H] values from SED fitting indicated by the best-probability solution (see Section 3.1) are systematically metal-rich 
with [Fe/H] $>$ $-$1.0, and show large (more than 1~dex) differences with our compiled [Fe/H]. 
Consequently, we suggest that most young clusters in M31 might be more metal-rich, than the results from \cite{per02} indicate. 
We anticipate high precision spectroscopic observations for an extensive sample of young clusters in the future to clarify 
the metallicity distribution of young clusters.

\subsection{Kinematics}

Of the 617 compiled star clusters with measured radial velocity, the majority of the clusters are in the range of $-700$ - 100~\kms. 
The mean value of radial velocity and velocity dispersion is $-295$~\kms and 163~\kms, respectively. 
This mean radial velocity is consistent with the known system velocity of M31, V$_{M31} = -301$~\kms \citep{van00}, 
which we adopt in this paper.  

In Figure 14, we present radial velocities of 617 star clusters, regardless of their ages, with respect to the system velocity 
of M31 against the projected distance along the major axis (black circles in left panels) and their velocity distribution (black histograms in right panels).  
We also divide the sample into star clusters which are located in different bins along the minor axis (i.e., different Y range) 
in order to inspect kinematical variation along the minor axis. 
In each left panel, a linear fit to the sample within $|X|$ = 10~kpc (solid line), passing through $X$ = 0 and V$_{r}-$ V$_{M31}$ = 0, 
along the major axis is also shown. 
The slope ($\alpha$) of the linear fit decreases as the distance along the minor axis increases. 
It is evident from the figures that most of the star clusters in M31 show a sign of coherent rotation around the minor axis. 
The slope ($\alpha$) of the linear fit in each bin along the minor axis is large and the velocity distributions have 
two peaks around the system velocity. 
This is in good agreement with previous results \citep[e.g.,][]{per02,lee08}. 
The rotation feature of the clusters at $|Y| <$ 3~kpc is more evident compared with the counterparts at  $|Y| >$ 3~kpc.

In Figure 14, we also present velocity distribution of old ($>$ 1~Gyr) clusters (red circles and red histograms). 
In both the whole old cluster sample and each subsample in different bins along the minor axis, a hint of rotation is also seen, 
but less prominent than in the whole sample. 
The clear rotation signature does not appear for the outermost clusters at $|Y| >$ 3~kpc. 
Furthermore, in each panel, large scatter around the linear fit is seen. 
These indicate that, while the pressure support plays a significant role, rotational kinematics are also important 
for the old cluster system in M31. 
In the case of old clusters at  $|Y| <$ 3~kpc, they might be dominated by a central large bulge showing moderate rotation 
rather than by a pressure supported halo (see \citealt{lee08} and references therein).

On the contrary, as shown in Figure 15, young ($\le$ 1~Gyr) star clusters show most striking feature of systematic rotation 
around the minor axis. 
The radial velocities of young clusters against the projected distance along the major axis show tight distribution with 
little scatters around the linear fit. 
This indicates that the system of young clusters is rotational supported in the M31 disk. 
The rotational velocity of young clusters at $|Y|$ $<$ 1~kpc is about 220~\kms. 

\cite{mor04} have suggested the existence of a cold thin-disk system of star clusters in M31. 
In order to further clarify the kinematics of young clusters, we attempt to select a subsystem of star clusters with thin-disk 
kinematics, therefore presumably associated with the disk of M31. 
Following \cite{mor04}, we considered a simple cold-disk kinematical model with zero-thickness.  
In this model, the position of a cluster in the disk is determined from its observed position on the sky. 
We use a rotation curve that is flat with V$_{cirular}$ = 250~\kms for $|R|$ $>$ 6.5~kpc and then falls linearly to zero at $X$ = 0 
(solid line in the bottom panel of Figure 16). 
We also adopt a distance to M31 of 785~kpc \citep{mcc05}, system radial velocity of $-301$~\kms, 
inclination of 77.7~deg \citep{van00}, and position angle of 37.7~deg \citep{dev58}. 
If star clusters lay on the disk, we could calculate their velocities using our assumed cold-disk model 
and disk rotation curve. 
Finally, we measured the residual velocity which is defined as the absolute value of the difference between the calculated 
velocity and the actual observed velocity (see middle panel of Figure 16).

We split the star clusters into two subgroups according to their residual velocity: 
thin-disk kinematics and non-thin-disk (i.e., bulge or halo) kinematics. 
We define thin-disk clusters as those with residual velocities less than 40~\kms (see dashed line in the middle panel 
of Figure 16) following the criterion of \cite{mor04}. 
About 90~\% of the star clusters with thin-disk kinematics of \cite{mor04} is recovered by our selection criteria. 
We divided 617 star clusters into 216 thin-disk clusters and 401 non-thin-disk clusters.
In the case of old clusters, only 29~\% of sample has thin-disk kinematics. 
On the other hand, 66~\% of the young clusters (blue filled circles in Figure 16) shows thin-disk kinematics. 
Therefore, most young clusters (filled circles in Figure 16) have thin-disk kinematic characteristics \citep[see also][]{fus05}. 
This argument is also supported by the spatial distribution of young clusters, which 
are preferentially located along the 10~kpc ring (red ellipse in the top panel of Figure 16) in the disk of M31 (see also Section 4.4).

\subsection{Spatial Distribution}

In the top panel of Figure 16, we compare the spatial distribution of young ($\le$ 1~Gyr, filled circles) and old ($>$ 1~Gyr, gray open circles) clusters 
in the plane of projected distances along the major ($X$) and minor ($Y$) axes. 
It is evident that old clusters are uniformly distributed all over M31 (from galaxy center to outermost disk regions), 
while many old clusters are concentrated near the central regions of M31 (i.e., bulge region). 
On the contrary, young clusters are lacking in the central regions of the galaxy and are evidently projected onto the disk 
between 5~kpc (inner dashed ellipse) and 18~kpc (outer dashed ellipse). 
Specifically, the spatial distribution of young clusters is well correlated with the well-known star-formation region associated with 
the 10~kpc ``ring of fire" in the M31 disk \citep{bri84, dam93, pag99}.

In Figure 17, we compare the spatial distribution of young clusters with respect to that of OB stars (top panel), 
UV SF regions (middle panel), and dust (bottom panel). 
Young clusters in different age ranges (age $<$ 100~Myr, 100~Myr $<$ age $<$ 400~Myr, and 400~Myr $<$ age $<$ 1~Gyr) are shown in different colors. 
We select O and B type stars (gray dots in the top panel) from $UBVRI$ photometric data of 
\cite{mas06} \citep[see][]{kan09}. 
We also consider 894 SF regions (gray contours in the middle panel) defined from the $GALEX$ FUV imaging \citep{kan09}. 
The spatial distribution of dust from a $Spitzer$ IRAC 8.0 $\micron$ non-stellar image \citep{bar06} is also presented (contours in the bottom panel). 
The 10~kpc ring is approximated by a red ellipse. 
Evidently, the spatial distribution of young clusters correlate well with that of OB stars, 
UV SF regions, and dust as well as with the 10~kpc ring structure. 
However, it is noteworthy that the OB stars and UV SF regions spread out to the outer parts of the M31 disk, while 
young clusters are mostly confined to the regions around 10~kpc.
There may be a selection effect favoring detection of UV SF regions and hot stars in the outer disk regions where extinction is less. 

Figure 18 presents the number histogram of star clusters against the distance from the center of M31. 
OB stars and UV SF regions are also shown for comparison. 
A noticeable feature is that old clusters show a very different distribution from those of young clusters, OB stars, and UV SF regions. 
Old clusters are more centrally concentrated within $\sim$10~kpc. 
While young clusters follow a similar distribution to OB stars and UV SF regions, young clusters show a peak around 10~kpc - 12~kpc. 
On the other hand, the distribution of OB stars and UV SF regions have an additional peak around 16~kpc. 
Consequently, we suggest that young clusters are closely correlated with OB stars and UV SF regions in their spatial distributions, 
although OB stars and UV SF regions show a more extended structure in the disk of M31.
On the other hand, \cite{azi11} presented 3,691 HII regions on the disk of M31. 
They also found a reasonable spatial correlation between the luminous ($L_{H_{\alpha}} > 10^{36} erg^{-1}$) 
HII regions and young clusters \citep[see Figure 9 of][]{azi11}.

\subsection{Young Clusters and Star Formation Ring Structure}

In Figure 19, we compare ages of young clusters (middle left panel) and UV SF regions (bottom left panel) against 
the de-projected distance from the center of M31. 
The overall young clusters distribution shows a single peak around 10 - 12~kpc. 
In addition to this main feature, a hint of small secondary peak is seen around 13 - 14~kpc.
While young clusters in various age ranges within 1~Gyr contribute to the main peak, the small secondary peak might 
be ascribed to younger clusters with ages less than 400~Myr (see inset of the middle left panel). 
On the contrary, UV SF regions show two distinct peaks: a highest peak at $\sim$16~kpc and a secondary peak around 11~kpc. 
The ages of most UV SF regions are younger than 400~Myr mostly due to our UV-based selection \citep{kan09}. 

If young clusters are the result of active star formation in the M31 disk due to the head-on collision by a satellite galaxy, 
the age range of the majority of young clusters might be relevant to establishing the epoch of that event. 
Although the age distribution of our sample is somewhat broad, the majority of young clusters is in the age range of 100 - 400~Myr. 
Thus, our results appear to be consistent with the prediction by \cite{blo06} of a collisional event with M32 about 210~Myr ago. 
\cite{cal09} claimed a possible spatial age variation among young clusters, considering models of  recent 
interaction between M32 and the M31 disk and outward propagation of star burst through the disk by expanding density waves \citep{gor06,blo06}. 
We find no clear evidence of radial trend of cluster ages \citep[see also][]{cal09}.

As shown in Figure 19 (middle right panel), most young clusters have masses between 10$^{3.5}$ and 10$^{4.5}~M_{\sun}$.
While there is no clear correlation between mass and radial distance, interestingly, young clusters located around 10 - 14~kpc 
have a wider mass range, compared with those in other regions. 
In this region, both high-mass ($>$ 10$^{4.5}~ M_{\sun}$) and low-mass ($<$ 10$^{3.5}~M_{\sun}$) young clusters are found. 
The UV SF regions by \cite{kan09} have a mass range wider than that of the young clusters and the majority have 
masses between 10$^{3}$ and 10$^{5.5}~M_{\sun}$. 
It is worth noting that more massive ($>$ 10$^{5.5}~M_{\sun}$) UV SF regions are also preferentially located at galactocentric 
distances around 10 - 14~kpc where massive young clusters are also found. 
Another group of massive UV SF regions is found around 5~kpc. 
On the contrary, low-mass UV SF regions are more numerous towards outer regions ($>$ 10~kpc) of the M31 disk, and form a peak around 16~kpc. 
Based on the rarity of young clusters and systematically low-mass (and younger) UV SF regions outside 16~kpc, 
we suggest that the outer part of the disk of M31 has an environment insufficient to trigger formation of massive star clusters. 
This is also supported by the HI and CO surveys which show higher gas density around the 10~kpc ring and lower in outer parts \citep[e.g.,][]{bri84,dam93,nie06}. 

Using $Spitzer$ MIPS images, \cite{gor06} suggested that the morphology of the dust in M31 is well represented 
by a composite of two logarithmic spiral arms and a 10~kpc circular star forming ring offset from the nucleus. 
Following the approach of \cite{gor06}, in Figure 20 we examine the spatial arrangement of UV SF regions and young clusters. 
In the upper panels, the black large circle is the 10~kpc star formation ring, offset from the center of M31 by (4.5~kpc, 1.0~kpc) 
and with a radius of 44~arcmin (10~kpc). 
The two gray spirals are simple logarithmic spirals adopted from \cite{gor06}. 
Most UV SF regions do not follow the two spiral arms and show deviations 
from the spiral pattern, while some UV SF regions in the inner disk (inside 10~kpc) appear to roughly follow two logarithmic spirals. 
The distribution of many UV SF regions is well fitted by a 10~kpc circle.  
The UV SF regions in the outer disk (outside 12~kpc) present several arcs rather than spiral arms pattern. 
The distribution of young clusters is more distinct and simple; most young clusters are 
around the 10~kpc circular ring, while two spirals trace a handful of young clusters inside the 10~kpc ring. 
To confirm the distribution of UV SF regions and young clusters around the 10~kpc ring, in the lower panel of Figure 20 
we present their spatial distribution in polar coordinates. 
The position angle is defined to be increasing clockwise starting from northeast (NE) in the upper panels of Figure 20. 
The most interesting feature is the wavy distribution of SF regions and young clusters fitted by a black solid curve, 
representing the $\sim$10~kpc ring, except for the region around 220~deg where the ring splits. 
This wavy line is characterized by parameters of the circle shown in the upper panels of Figure 20. 
This wavy feature is more distinct in the case of young clusters (colored filled circles). 
Consequently, we confirm that the distribution of young clusters and UV SF regions in the M31 disk is consistent with 
a circular star formation ring with radius $\sim$ 10~kpc in combination with simple logarithmic spirals.

Another thing to be noted in Figure 20 is an asymmetric number distribution of young clusters with position angle. 
The number (125) of young clusters in the southern disk of M31 (from 180 to 360~degree) is larger than that (57) of young clusters 
in the northern disk (from 0 to 180~degree). 
\cite{fan08} noted that the highly extincted star clusters with $E(B-V) >$ 0.4~mag are preferentially located 
on the northwestern (NW) side \citep[see also Figure 17 of][]{cal11}. 
If the difference is entirely due to the reddening being higher on the NW half, the frequency and distribution of young clusters 
in the northern disk based on currently available catalogs remains to be updated from extensive and deeper observations in future.

\subsection{Ring Splitting Region and Compact Star Clusters}

A prominent feature in the southern part of the M31 disk is the region of ring split. 
In the distribution of both \cite{kan09} UV SF regions and our young clusters, a hole that matches the observed split in the ring near M32 
is seen around $X$ = $-$8~kpc and $Y$ = 8~kpc (upper panels in Figure 20) and position angle $\sim$160 - 270~deg (lower panel in Figure 20). 
This was discovered by \cite{gor06} from the IR-emitting dust distribution, and they suggested that the split of the ring 
in the form of a hole is caused by a passage of M32 through the M31 disk \citep[see also][]{blo06}. 
Many UV SF regions and young clusters are located outside of the ring splitting. 
As shown in the lower panel of Figure 20 (histogram of position angle), young clusters and SF regions show distinct peaks around 180 and 240~deg. 
These peaks appear to be constituted mostly of clusters younger than 400~Myr. 

Recently, based on high-resolution SUBARU Suprime-Cam images, \cite{van09} carried out a survey of compact star clusters 
in the southwestern part of the M31 disk including the ring splitting region. 
The apparent size of these compact clusters is less than 3~arcsec and for most sample it is smaller ($<$ 2.5~pc) than typical MW GCs. 
They estimated ages and masses of 238 high-probability star clusters based on the $UBVRI$ photometry from the LGGS images. 
These star clusters are mainly selected by specifying a lower limit of half-light radius ($r_{h} \gtrsim 0.15$~arcsec or $\gtrsim 0.6$~pc, see \cite{van09} for the details). 
The majority of their compact clusters are young objects, with ages less than 1~Gyr (186 of 238) and a peak around 70~Myr. 
They span a mass range of 10$^{2.0}$ - 10$^{4.3}~M_{\sun}$ peaking at $\sim4\times10^{3}~M_{\sun}$. 

In Figure 21, we present age and mass distribution of our young clusters (black filled and open circles) and 186 compact star 
clusters (gray filled circles) younger than 1~Gyr from \cite{van09}. 
While the overall distribution of our young clusters is consistent with that of \cite{van09} compact star clusters, 
it is obvious that \cite{van09} detected more objects in the lower mass ($<$ 10$^{3.5}~M_{\sun}$) 
and younger age ($<$ 100~Myr) ranges. 
About half (95) of their sample clusters are younger than 100~Myr. 
In their sample, there is a lack of relatively massive ($>$ 10$^{4}~M_{\sun}$) compact clusters with ages 400~Myr - 1~Gyr. 
Young clusters included in our compiled catalog show systematically higher mass (see mass histogram of Figure 21) than the 
cluster sample of \cite{van09}, probably due to the brighter limit of cluster selection used in previous literature. 

In Figure 22, we present the spatial distribution of 186 compact star clusters (filled circles) with ages younger than 1~Gyr 
from \cite{van09} along with our young clusters (open circles). 
The clusters are divided into two age groups: younger than 100~Myr (blue) and 100~Myr - 1~Gyr (red). 
We also overplot the distributions of dust from $Spitzer$ IRAC 8.0~$\micron$ image (gray contours) and UV SF regions (orange contours). 
The distribution of clusters younger than 100~Myr follows well that of dust and UV SF regions. 
Furthermore, as noted by \cite{van09}, two clumps of young clusters are found at ($X$, $Y$) $\sim$ ($-$10.2~kpc, $-$2.2~kpc) 
and ($-$10.5~kpc, $-$0.5~kpc). 
On the other hand, relatively older ($>$ 100~Myr) clusters are widely distributed over the area. 
Older ($>$ 100~Myr) clusters are preferentially located in the gap of the ring splitting area 
(e.g., $-$9.5~kpc $<$ $X$ $<$ $-$7~kpc and $-$2.5~kpc $<$ $Y$ $<$ $-$0.5~kpc) where dust and UV SF regions lack. 
Regarding the scenario whereby the passage of M32 through the M31 disk triggered a burst of star formation \citep{gor06}, 
we speculate that older ($>$ 100~Myr) clusters in the ring splitting hole were 
formed at the epoch of the first passage of M32, while formation of younger ($<$ 100~Myr) clusters around the hole region 
might be induced by later shock propagation.

\section{Summary and Conclusion}

We constructed a comprehensive star cluster catalog which contains 700 M31 star clusters compiled from 
RBC v4 \citep{gal04}, \cite{cal09,cal11}, and \cite{pea10}. 
We detected 418 and 257 clusters in $GALEX$ NUV and FUV, respectively, above flux limits of 23.7 and 23.6 ABmags and measured their UV magnitudes. 
Our catalog includes photometry in up to 16 passbands ranging from FUV to NIR as well as ancillary 
information such as reddening, metallicity, and radial velocities. 
Our merged catalog is the most extensive and updated one of UV photometry for M31 star clusters, 
superseding our previous UV catalog \citep{rey07}. 
We estimated ages and masses of star clusters by multi-band SED fitting; the UV photometry 
enables more accurate age estimation of young clusters. 

We also extracted a sample of 182 young clusters with ages less than 1~Gyr consisting of 173 clusters with 
our age estimation and 9 clusters from \cite{cal09}. 
Our estimated ages and masses of young clusters are in good agreement with previous literature \citep[e.g.,][]{bea04,cal09,van09,per10}. 
We examined the properties of young clusters such as age and mass distribution, metallicity, kinematics, and spatial distribution 
which provide unique probes of the star formation history of the disk of M31, and the prominent 10~kpc ring structure in particular. 
The mean age and mass of the young clusters are about 300~Myr and 10$^{4}~M_{\sun}$, respectively. 
The mass range of our young clusters in M31 is similar to that of massive clusters found in the LMC and is in between those of 
Galactic young clusters and old GCs \citep{bea04,fus05,cal09}. 
Since most low-mass ($<$ 10$^{5}~M_{\sun}$) clusters are young objects, we consider they may be disrupted within a few Gyrs \citep{lam05}. 

The [Fe/H] values of young clusters included in our catalog, which are mostly from \cite{per02}, 
are systematically lower (by more than 1~dex) than those derived from high-quality spectroscopic data or inferred from our SED fitting. 
While high precision spectroscopic observations for an extensive sample of young clusters are anticipated, 
we suggest that most young clusters in M31 might have  moderately enhanced metallicity (i.e., [Fe/H] $>$ $-$1.0). 
Such a high metallicity of the M31 disk and associated young clusters is consistent with the general result that 
spiral's stellar disk is metal-rich \citep{bel00}, when not experiencing substantial gas infall or outflow \citep[e.g.,][]{dal07}. 

By comparing radial velocities of star clusters with a cold disk model, we selected a subsystem of star clusters with 
thin-disk kinematics, associated with the disk of M31. 
We confirm that most of the young clusters show systematic rotation around the minor axis and 
are kinematically associated with the thin-disk of M31. 
The majority of young clusters is located between 5 and 18~kpc from the center of M31 and shows a distinct peak around 10 - 12~kpc. 
The distribution of young clusters is closely correlated with that of other tracers of disk structure (OB stars, UV SF regions, and dust). 

Considering their kinematical properties and spatial distribution, young clusters are well correlated with 
the 10~kpc ring structure in M31. 
This structure, known as the ring of fire or star-formation ring, is off-centered from 
the galaxy nucleus (\citealt{blo06} and references therein). 
\cite{blo06} discovered an inner dust ring, offset from the center of M31, and suggested that the two rings 
originated from a recent passage of a satellite. 
By numerical simulations, \cite{blo06} inferred that the star formation ring structure results 
from a head-on collision through the center of the disk of M31 by a companion satellite galaxy. 
This event can produce density wave rings which triggered massive star formation at the peak of the wave. 
They postulated a recent (about 210~Myr ago) interaction between M32 and the M31 disk. 
We confirm that the spatial distribution of young clusters in the M31 disk is well re-presented by a circular 10~kpc star formation ring. 
Although the age distribution of our sample is somewhat broad, the majority of young clusters is in the age range of 100 - 400~Myr, 
which appear to be consistent with the prediction of \cite{blo06}. 
Therefore, we speculate that a large fraction of young clusters found in the M31 disk might have formed during the 
recent interaction between a satellite galaxy and the M31 disk, in which the star formation ring also originated. 
In particular, there is a split of the ring structure in the southern part of the M31 disk which corresponds to a gap in both 
IR contours and UV SF regions. 
Young clusters also show concentration outside the ring splitting and, furthermore, most of them have systematically 
younger ($<$ 100~Myr) ages. 
Some young clusters in this region might derive from another interaction with a satellite galaxy, related 
with the Southern Stream emerging from the southwest disk of M31 \citep{lee08}.

Within the context of merger history of M31, it is not unreasonable that the star formation ring in M31 
has been shaped by a recent collision of satellite accompanied by higher star formation rate than that of the MW \citep{ren05,yin09}. 
Due to a recent major perturbation of the M31 disk, formation of significant young stellar populations and 
massive young clusters is expected. 
Low-level star formation (e.g., in quiescent galactic disks) tends to produce few, if any, massive young clusters. 
It is worth noting that M31 appears to be representative of the typical population of local spiral galaxies 
showing evidence of merging in the formation and evolution history \citep[e.g.,][]{ham07}. 
On the other hand, the MW is a rather quiescent galaxy without any major interaction over the past few billion years.  
In this case, the MW disk may have evolved with a secular pattern \citep[e.g., smooth gas accretion or infall;][]{cro06,ham07} 
without any violent merging event. 
This kind of quiescent environment of the Galactic disk can support the nonexistence of the populous massive young clusters found in M31.
On the other hand, although head-on collisions between galaxies are rare \citep[see][]{mad09}, 
M31 serves as an important local template, to understand more distant collisional ring galaxies (\citealt{moi09} and references therein).

\acknowledgments
We are grateful to P. Barmby for kindly providing the $Spitzer$ IRAC image. 
The authors would like to thank the anonymous referee for thoughtful comments that helped to improved this paper. 
The $GALEX$ data presented in this paper were obtained from the Multimission Archive at the Space Telescope Science Institute (MAST). 
This research was supported by Basic Science Research Program through the National Research Foundation of Korea (NRF) 
funded by the Ministry of Education, Science and Technology (No. 2009-0070263). 
Support for this work was also provided by the NRF of Korea to the Center for Galaxy Evolution Research.
$GALEX$ (Galaxy Evolution Explorer) is a NASA Small Explorer, launched in April 2003.
We gratefully acknowledge NASA's support for construction, operation, and science analysis of the $GALEX$ mission, developed in cooperation with the Centre National d'Etudes Spatiales of France and the Korean Ministry of Science and Technology.


\begin{figure}
\includegraphics[width=80mm]{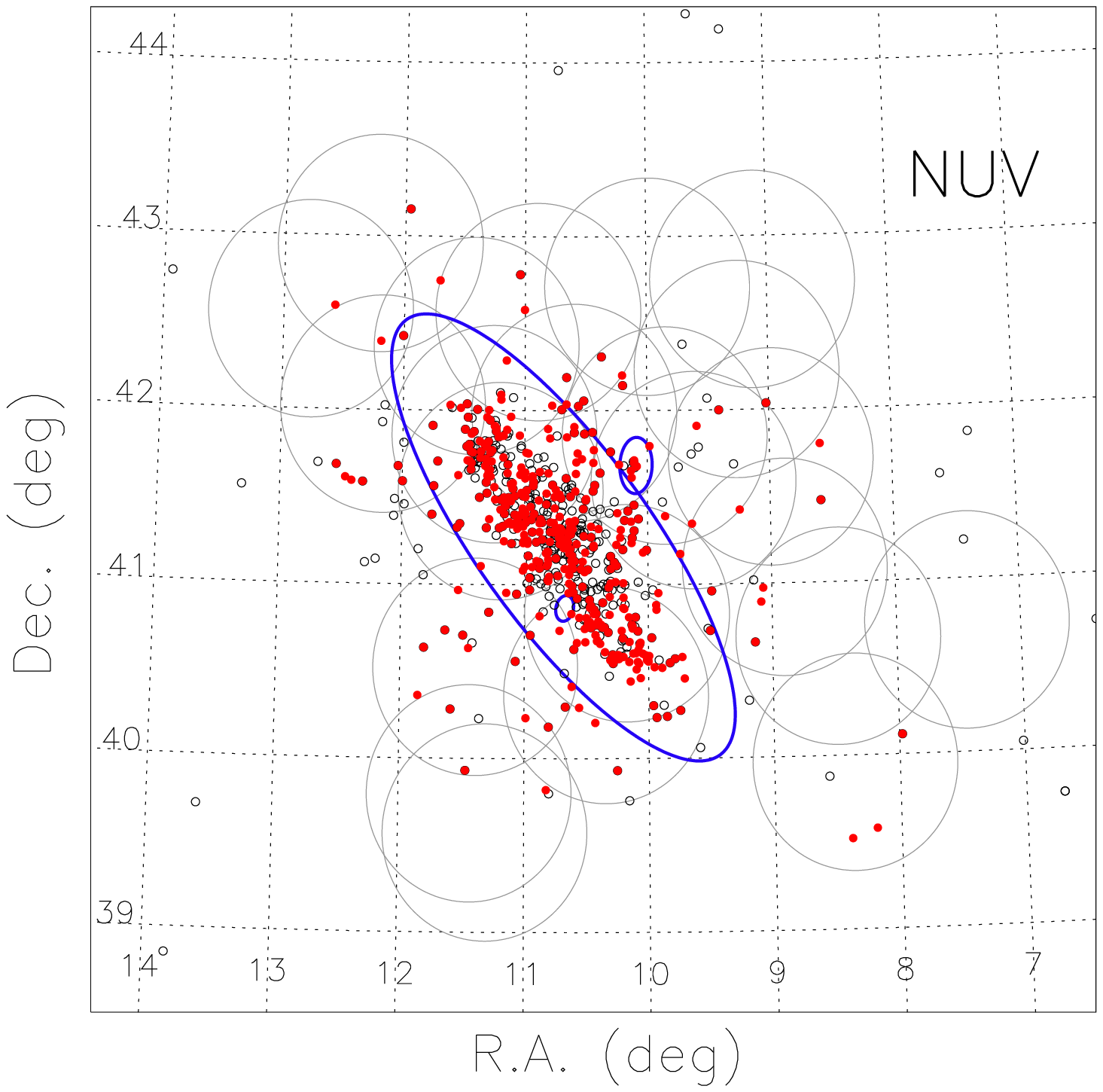}\\
\includegraphics[width=80mm]{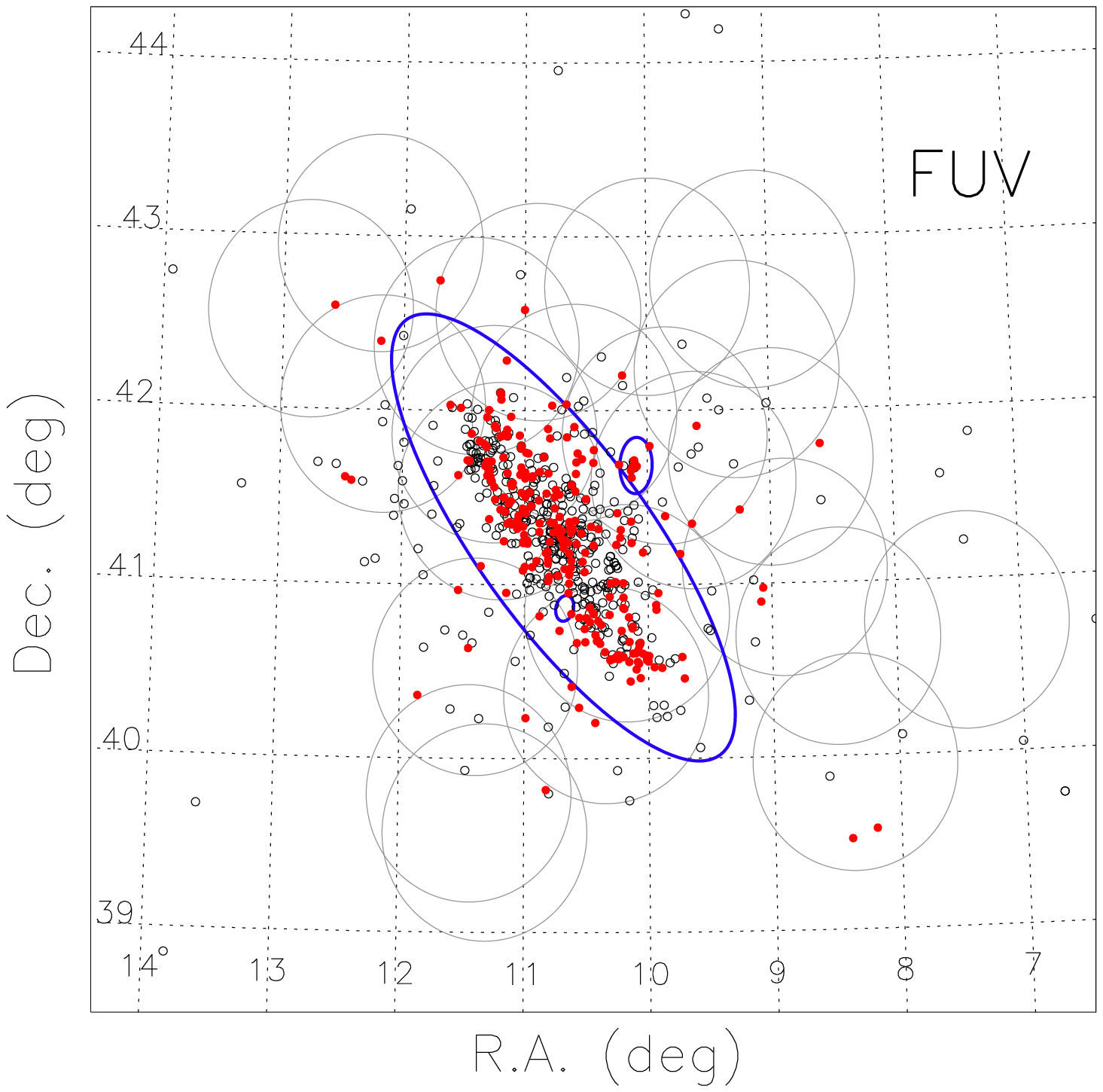}
\caption{
Spatial distribution on the sky of star clusters detected in $GALEX$ NUV (\textit{top panel}) and FUV (\textit{bottom panel}) fields. 
Of the 700 star clusters, red filled circles are UV detected ones and black open circles are those not detected in UV. 
The large blue ellipse is M31 and two smaller ellipses are NGC 205 (\textit{larger ellipse}) and M32 (\textit{smaller ellipse}) with the $D_{25}$ isophotes \citep{kar04}. 
Gray circles are 23 $GALEX$ fields.
\label{fig1}}
\end{figure}
\clearpage

\begin{figure}
\includegraphics[width=80mm]{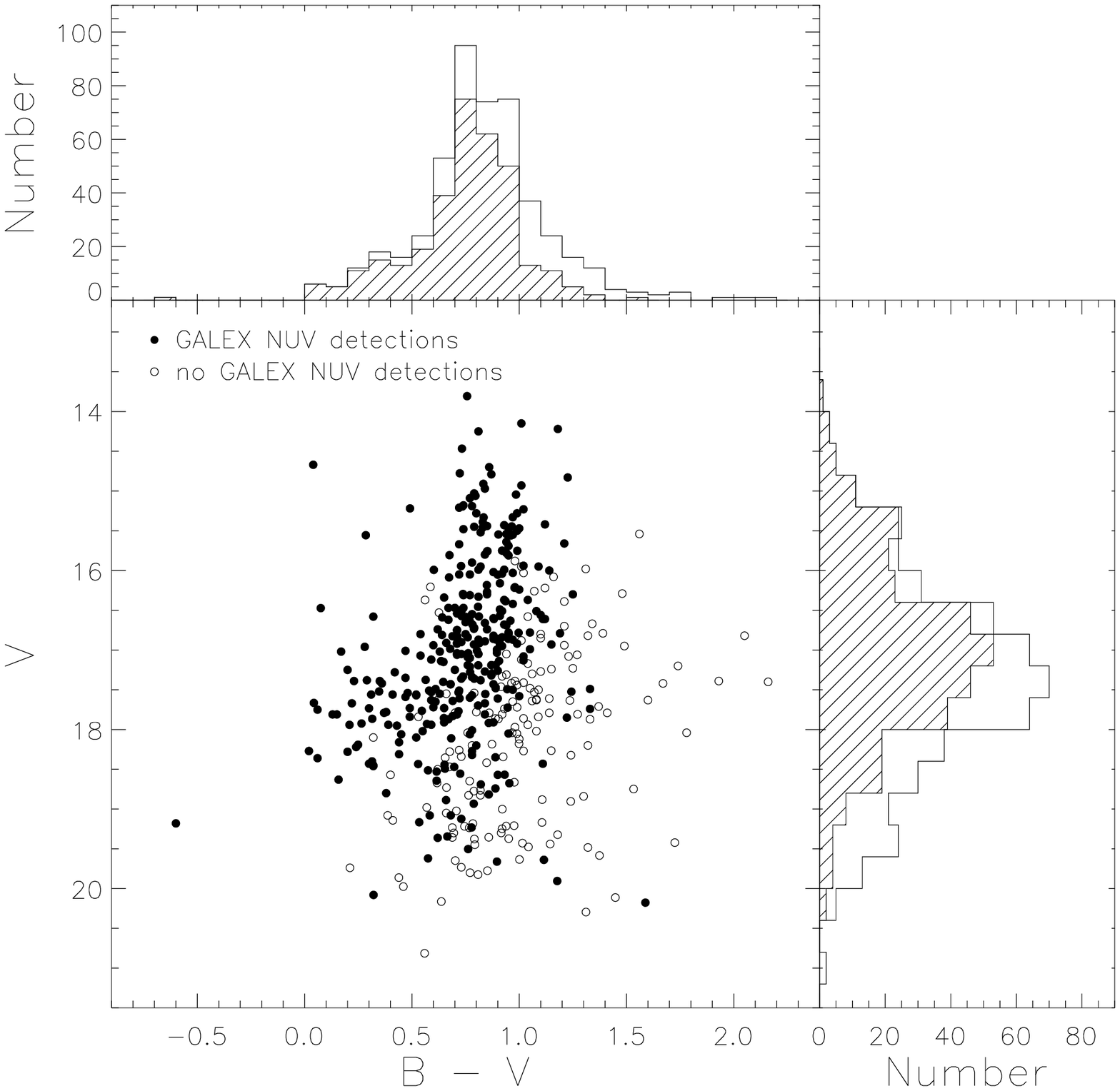}\\
\includegraphics[width=80mm]{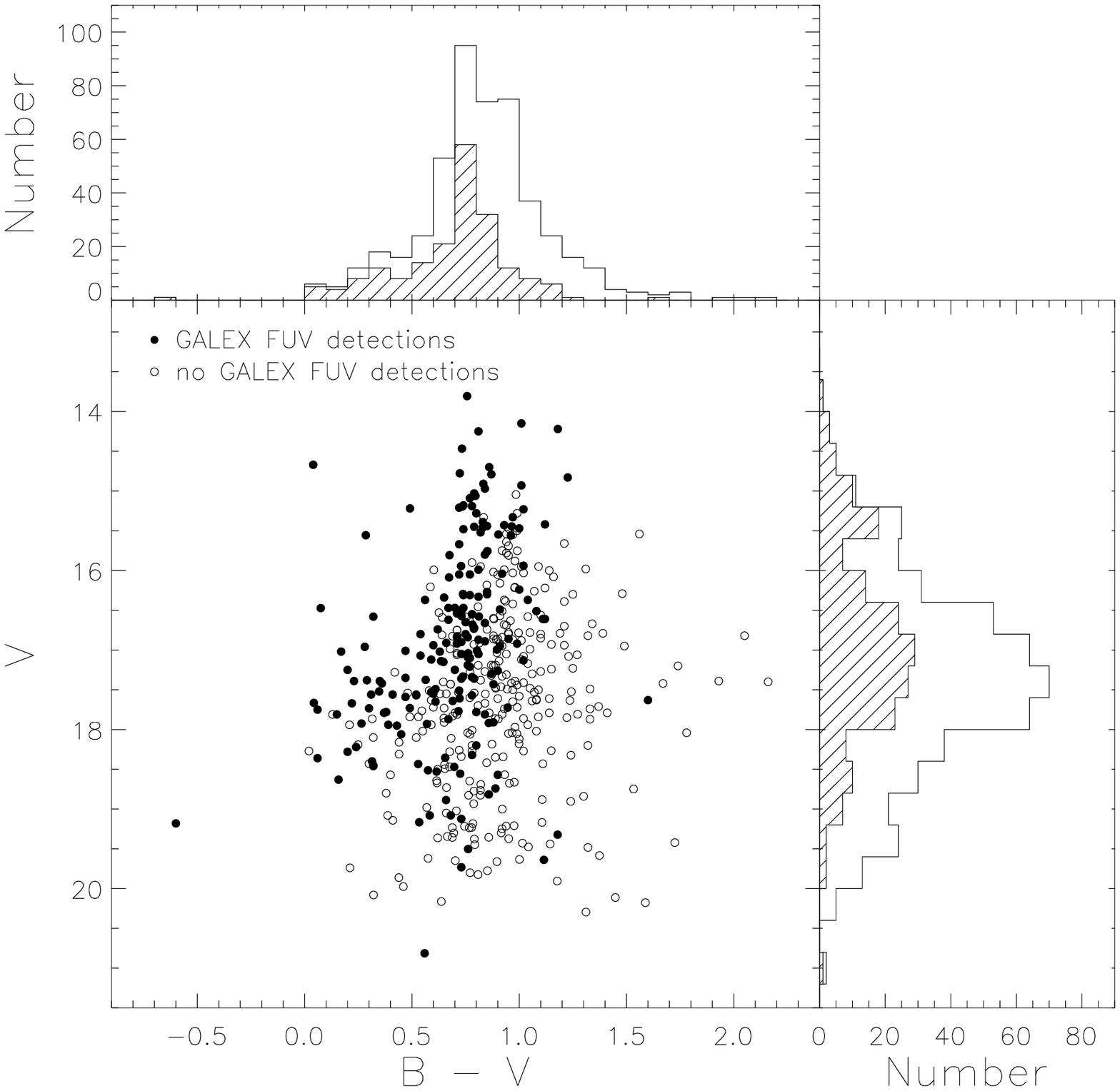}
\caption{
$GALEX$ detection rates in NUV (\textit{top panel}) and FUV (\textit{bottom panel}). 
Of the 484 clusters with both B and V data, 328 (about 68~\%) and 191 (about 39~\%) clusters are detected in the $GALEX$ NUV and FUV, respectively. 
\label{fig2}}
\end{figure}
\clearpage

\begin{figure}
\includegraphics[width=120mm]{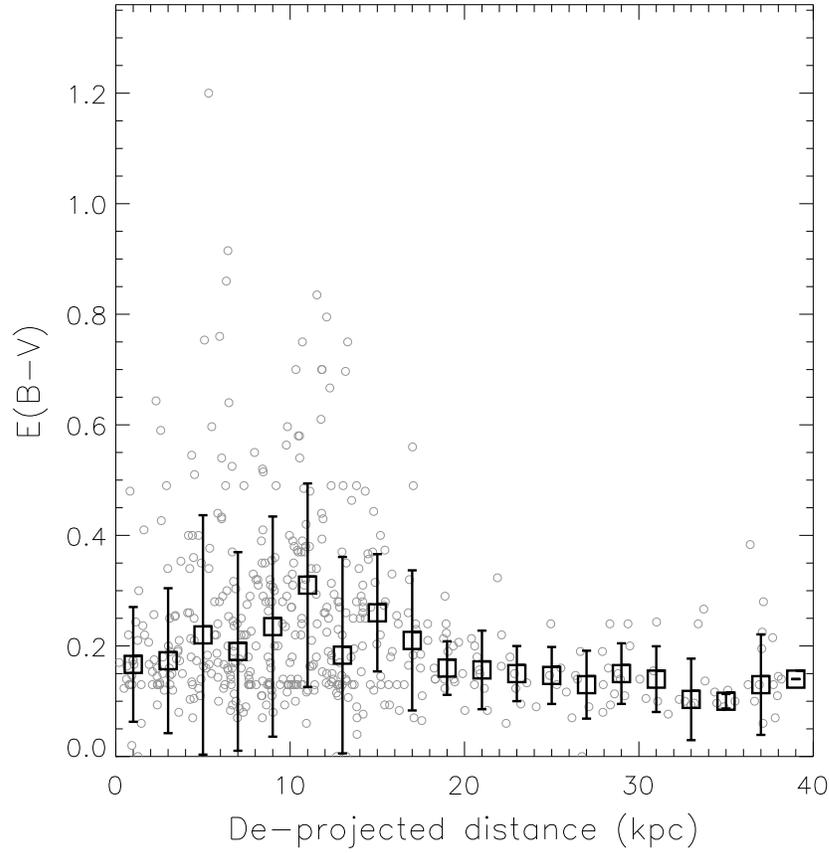}
\caption{
Distribution of our compiled reddening values of 555 star clusters (open circles) against their de-projected distances 
from the center of M31. 
Squares and error bars are median reddening values and standard deviations of star clusters located within annuli at 
every 2 kpc from the center of M31. 
The NW half of the disk has higher $E(B-V)$ on average than the SE region but the difference is less than respective standard deviations. 
\label{fig3}}
\end{figure}
\clearpage

\begin{figure}
\includegraphics[width=140mm]{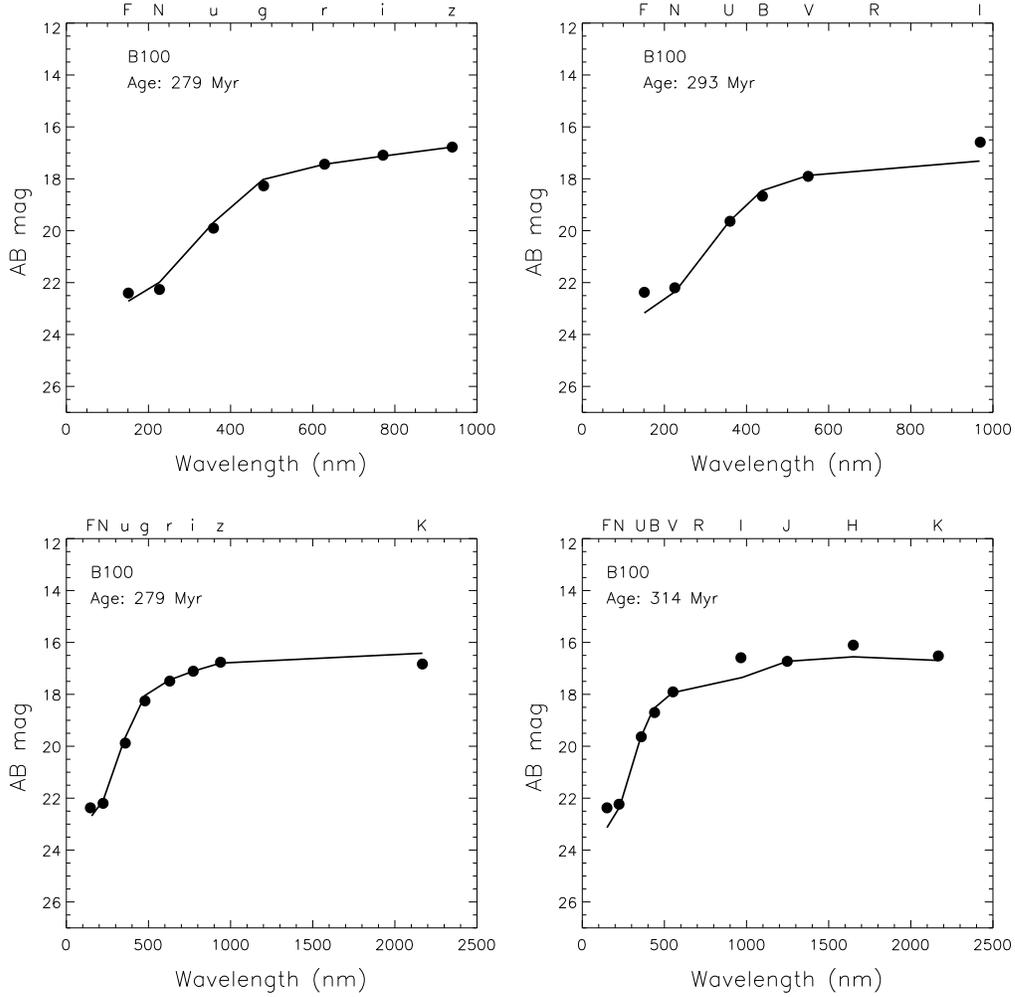}
\caption{
Examples of SED fitting for FN$ugriz$, FN$UBVRI$, FN$ugrizK$, and FN$UBVRIJHK$ photometry. 
Filled circles are photometric measurements at different passbands in AB mag system. 
The object name and the estimated age are indicated in the left upper corner of each panel. 
Reddening values are adopted from our compiled catalog. 
\label{fig4}}
\end{figure}
\clearpage
 
\begin{figure}
\includegraphics[width=140mm]{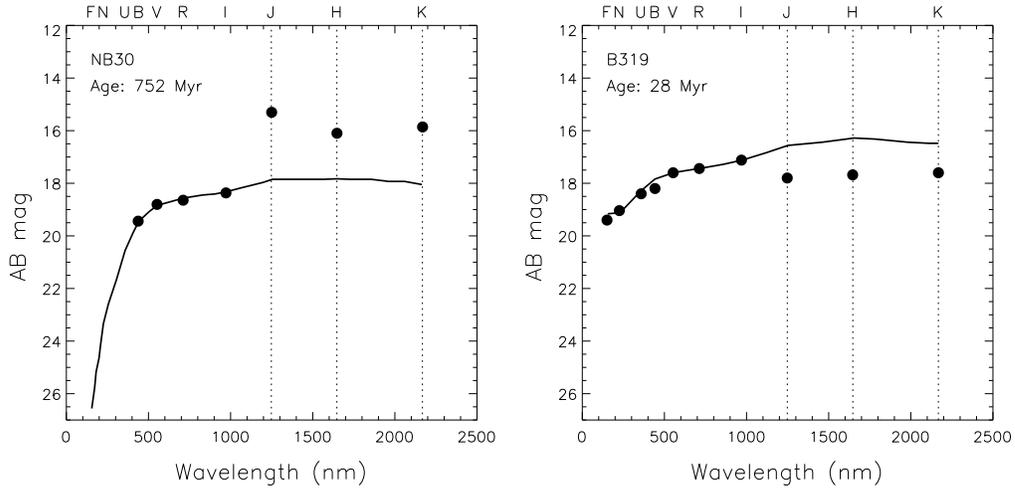}
\caption{
Examples of SED fitting for FUV, NUV, $UBVRI$, where the $JHK$ photometry is inconsistent. 
Filled circles are photometric measurements at different passbands in AB mag system. 
Vertical dotted lines indicate $JHK$ bands. 
The object name and the estimated age are indicated in the left upper corner of each panel. 
Reddening values are adopted from our compiled catalog. 
The photometry in $JHK$ bands shows large discrepancies by $>$ 1~mag from UV and optical bands regardless of the adopted best-fit model. 
\label{fig5}}
\end{figure}
\clearpage
 
\begin{figure}
\includegraphics[width=95mm]{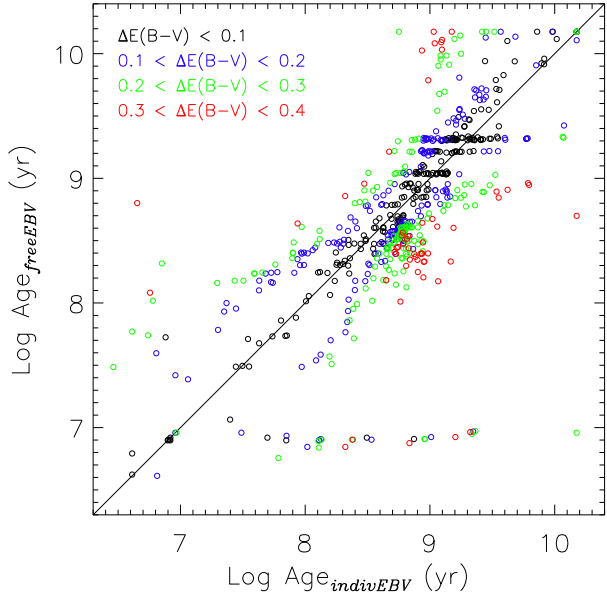}
\caption{
Comparison of ages derived from SED-fitting (1) imposing the E(B-V) values from 
literature ($indivEBV$, X axis) and (2) treating E(B-V) as a free parameter ($freeEBV$, Y axis). 
The symbols are color-coded according to the difference between $indivEBV$ and $freeEBV$ in the 
four cases: black ($\Delta E(B-V) < 0.1$), blue ($0.1 < \Delta E(B-V) < 0.2$), 
green ($0.2 < \Delta E(B-V) < 0.3$), and red ($0.3 < \Delta E(B-V) < 0.4$).
The example shows results from the FN$ugriz$ dataset, analyzed with Z=0.02 metallicity models.
\label{fig6}}
\end{figure}
\clearpage

\begin{figure}
\includegraphics[width=100mm]{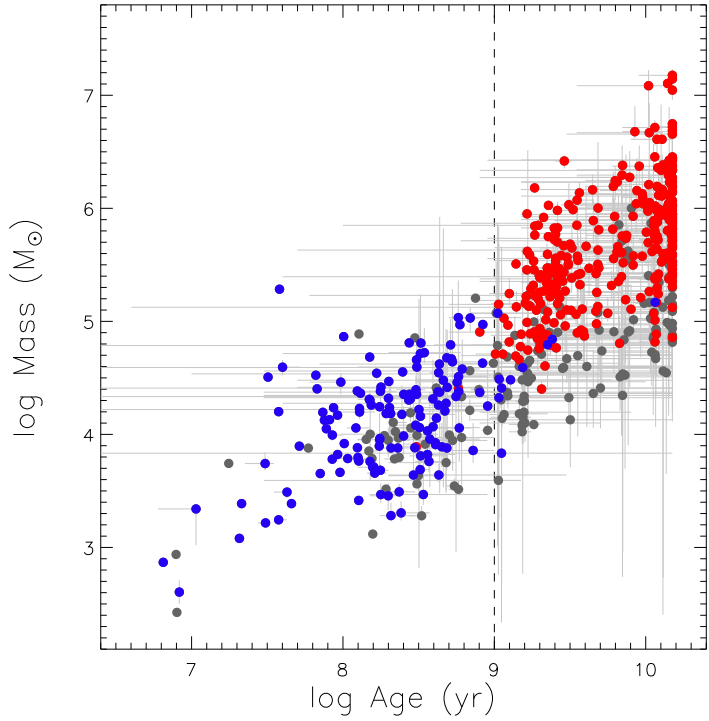}
\caption{
Distribution of estimated age and mass of star clusters in M31 from SED fitting. 
Blue filled circles are star clusters younger than 1~Gyr and red filled circles are those older than 1~Gyr 
according to the age estimated by \cite{cal09,cal11}. 
Gray filled circles are clusters with no age estimates from \cite{cal09,cal11}. 
Vertical dashed line indicates 1~Gyr.
Error bars are uncertainties in the estimated ages and masses from $\chi^2$ contours which are 
equal to minimum $(\chi^2)+1$.
\label{fig7}}
\end{figure}
\clearpage

\begin{figure}
\includegraphics[width=130mm]{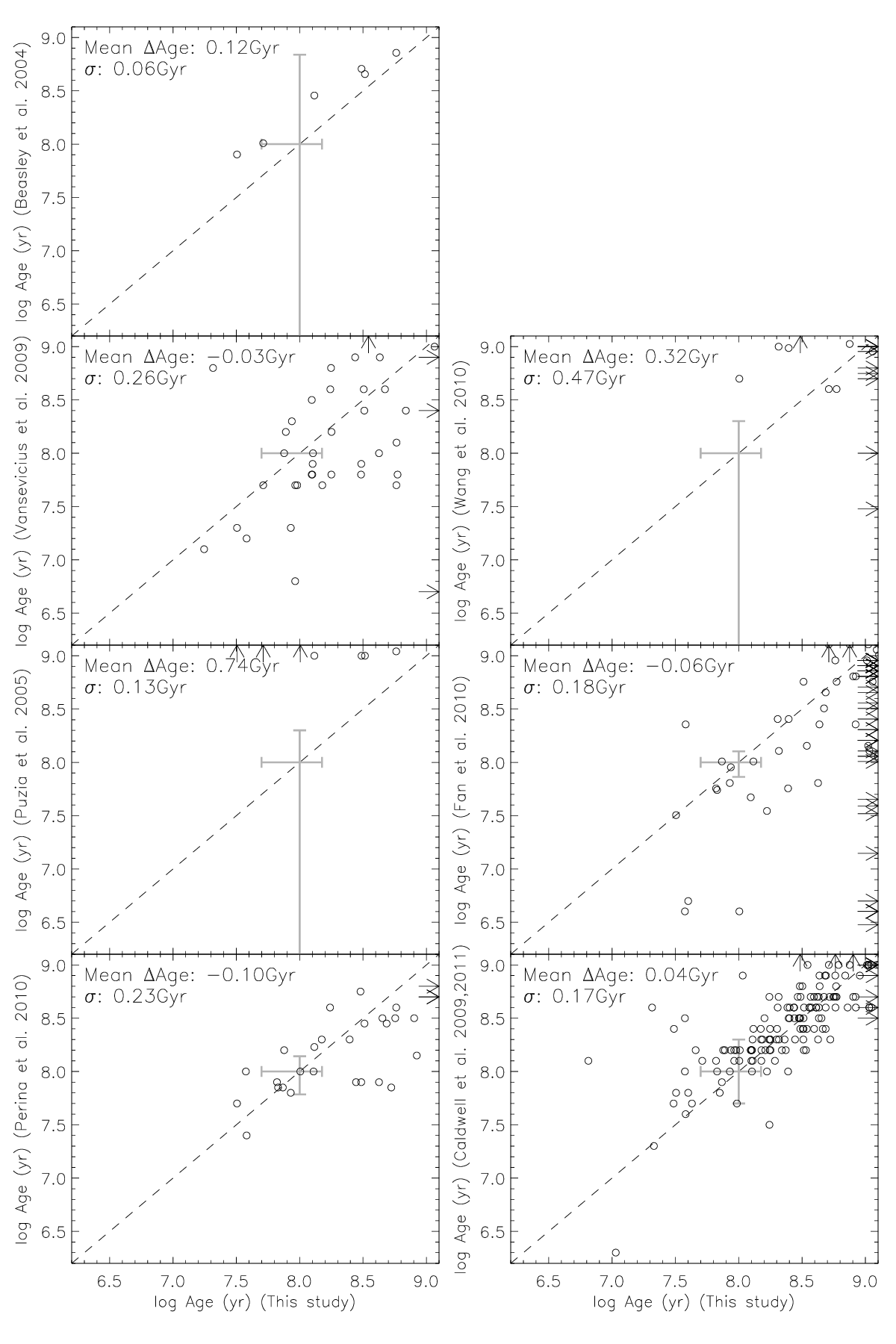}
\caption{
Comparison between age from our analysis and other studies for young clusters: \cite{bea04},  
\cite{van09}, \cite{puz05}, \cite{per10}, \cite{wan10}, \cite{fan10}, and \cite{cal09,cal11}. 
In each plot, the mean value of the age difference (other study minus ours) is given, with its standard deviation ($\sigma$).
Error bars show median errors of cluster ages.
\label{fig8}}
\end{figure}
\clearpage

\begin{figure}
\includegraphics[width=140mm]{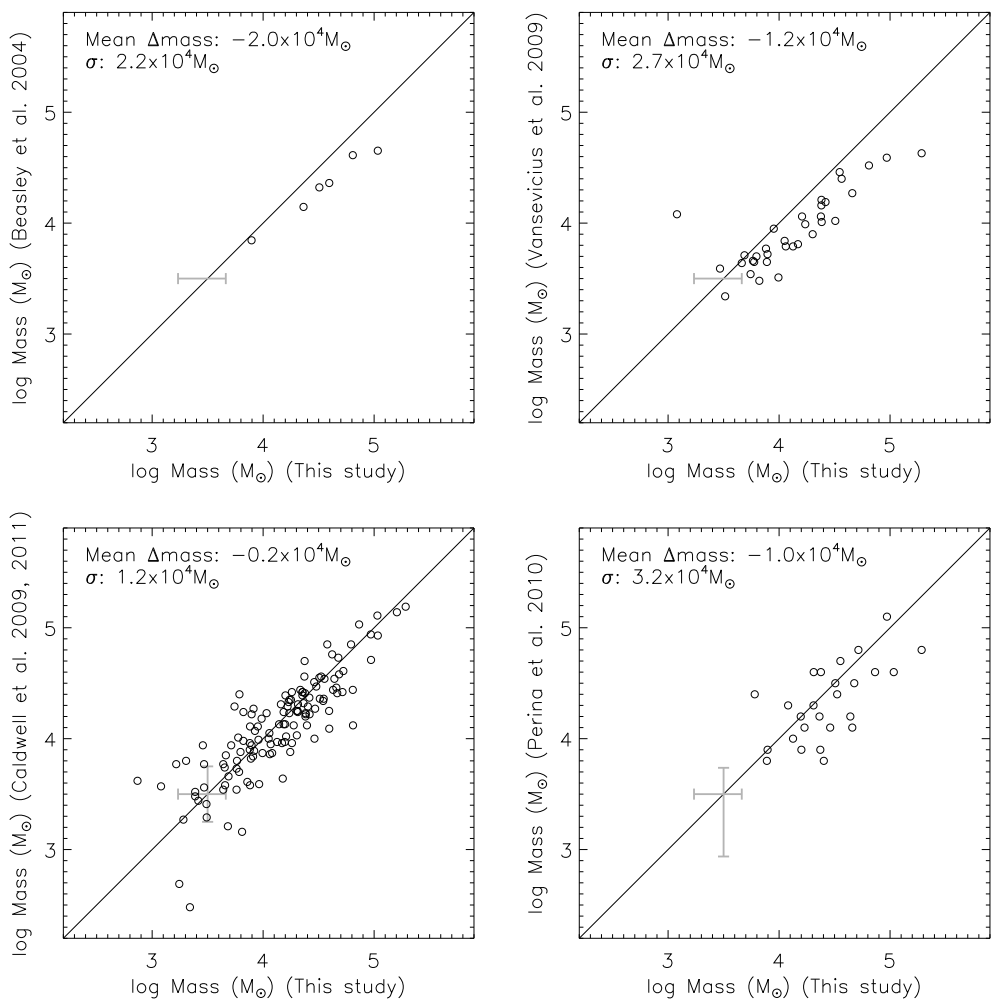}
\caption{
Comparison between cluster masses derived in this study and others: 
\cite{bea04}, \cite{van09}, \cite{cal09,cal11}, and \cite{per10}. 
In each plot, the mean value of mass difference (other study minus ours) is given with the standard deviation ($\sigma$).
Error bars show median errors of cluster masses.
\label{fig9}}
\end{figure}
\clearpage

\begin{figure}
\includegraphics[width=140mm]{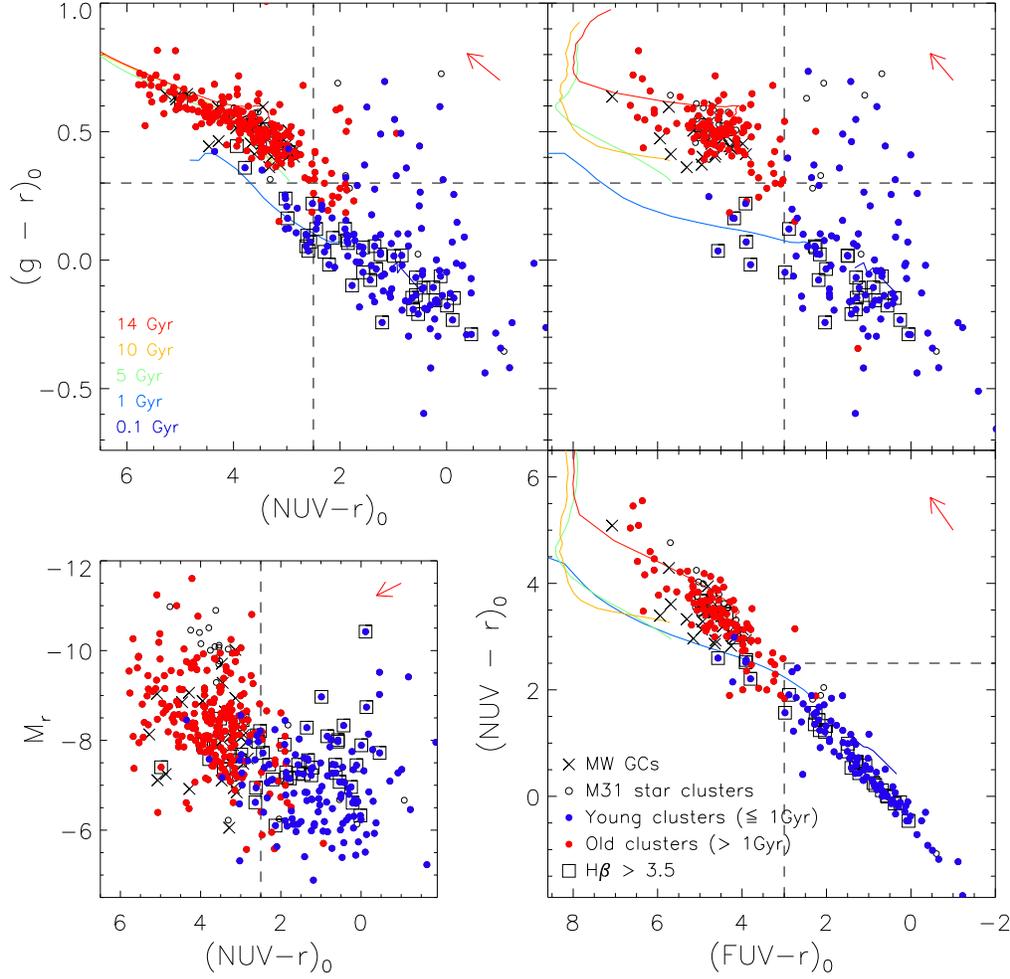}
\caption{
UV and optical color-color and color-magnitude diagrams of M31 star clusters. 
Blue and red filled circles are young ($\le$ 1~Gyr) and old ($>$ 1~Gyr) clusters, respectively. 
Open circles are objects with no age estimates from our analysis. 
Open squares are clusters with H$\beta$ spectral index larger than 3.5~\AA. 
MW GCs are plotted with crosses. 
Solid curves are Yonsei evolutionary population models in the age range 0.1 - 14~Gyr. 
Horizontal and vertical dashed lines indicate reference values for dividing young and old clusters; 
(FUV$-r$)$_0$ = 3.0, (NUV$-r$)$_0$ = 2.5, and $(g-r)_0$ = 0.3. 
The arrow indicates a reddening vector for $E(B-V)=0.10$~mag.
\label{fig10}}
\end{figure}
\clearpage

\begin{figure}
\includegraphics[width=140mm]{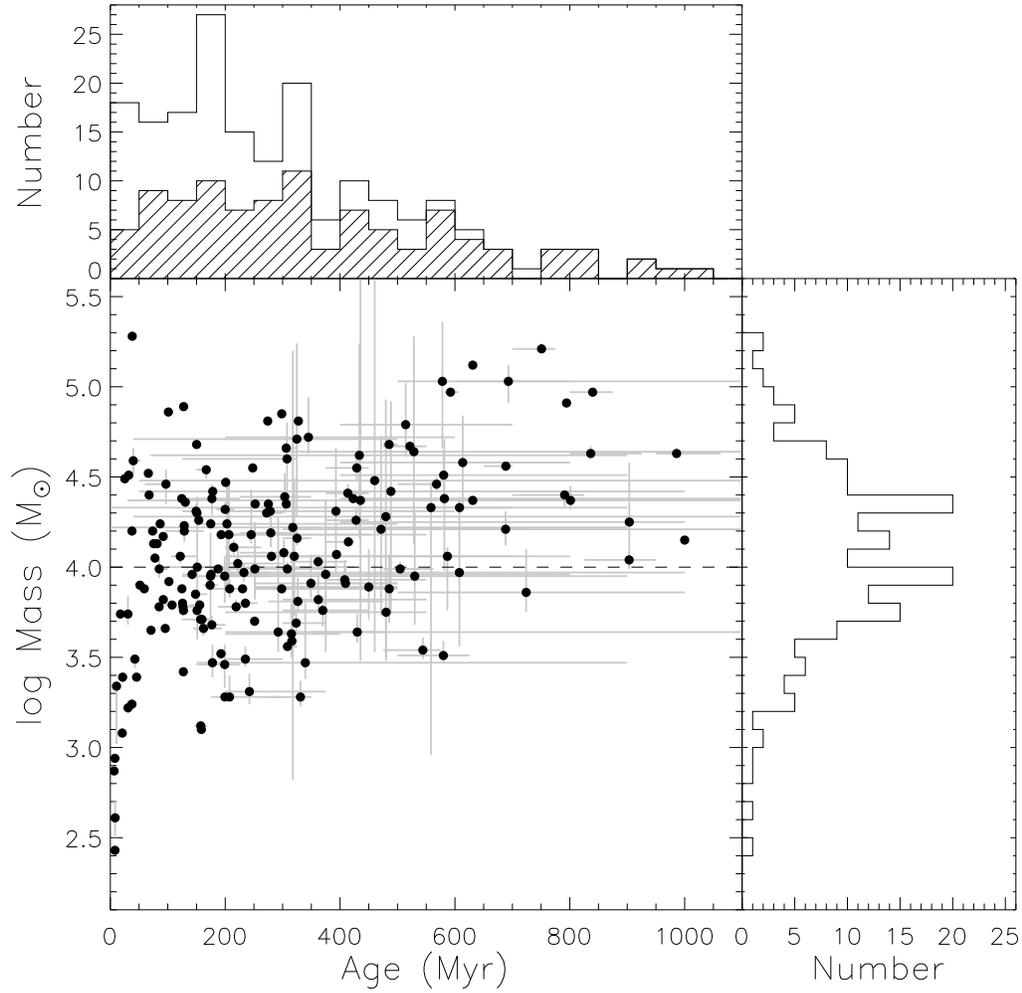}
\caption{
Age and mass distribution of 182 young clusters in M31. 
The hatched histogram is for clusters more massive than $10^4~M_{\sun}$.
Error bars are uncertainties in the estimated ages and masses from $\chi^2$ contours which are 
equal to minimum $(\chi^2)+1$.
\label{fig11}}
\end{figure}
\clearpage

\begin{figure}
\includegraphics[width=80mm]{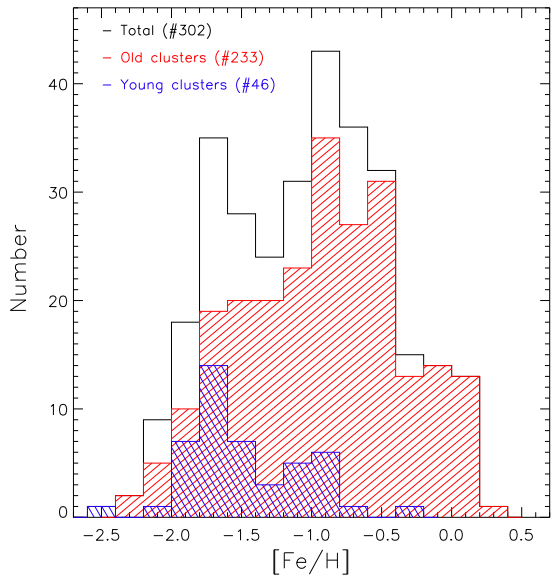}
\caption{
The metallicity distribution of the star clusters. 
Red and blue hatched histograms are for clusters older and younger than 1~Gyr, respectively. 
The black histogram is the distribution of the total cluster sample.
\label{fig12}}
\end{figure}
\clearpage

\begin{figure}
\includegraphics[width=80mm]{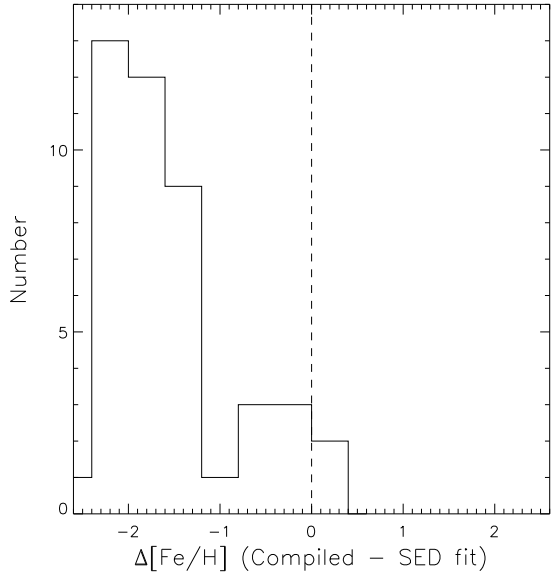}
\caption{
Metallicity differences of young clusters between our compiled values and values from our SED fitting.
\label{fig13}}
\end{figure}
\clearpage

\begin{figure}
\includegraphics[width=140mm]{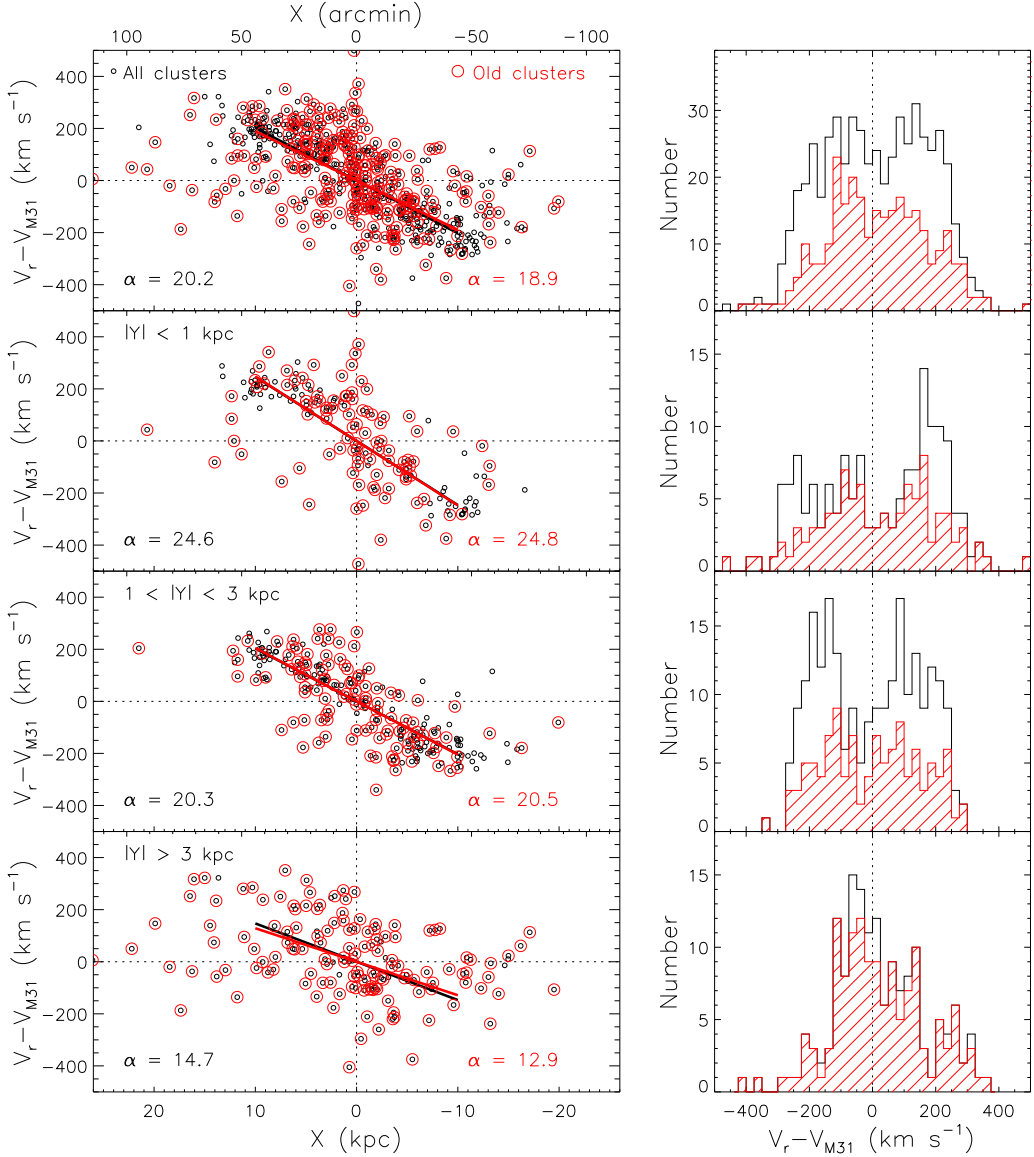}
\caption{
Radial velocities of 617 star clusters with respect to the M31 system velocity against the projected distance 
along the major axis (\textit{black circles in left panels}) and their velocity distribution (\textit{black histograms in right panels}). 
Red circles and hatched histograms are for old ($>$ 1~Gyr) clusters.
Solid lines in left panels are the linear fits to the sample within $|X|$ = 10~kpc and $\alpha$ is the slope of the fit. 
Top panels show the star clusters over the whole region. 
Other panels are for the star clusters at different $|Y|$ ranges. 
\label{fig14}}
\end{figure}
\clearpage

\begin{figure}
\includegraphics[width=140mm]{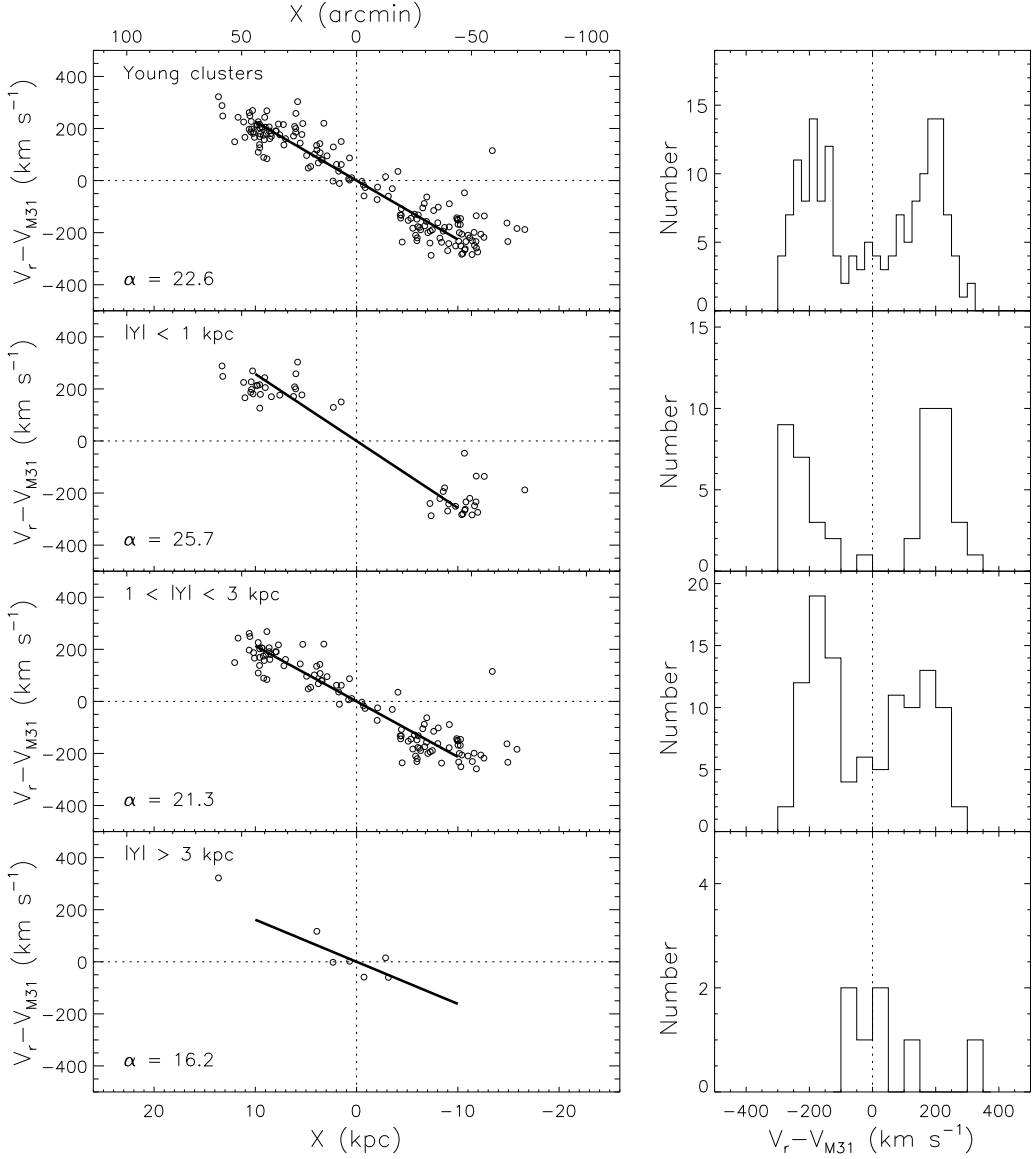}
\caption{
Same as Figure 14, but for young ($\le$ 1~Gyr) clusters.
\label{fig15}}
\end{figure}
\clearpage

\begin{figure}
\includegraphics[width=110mm]{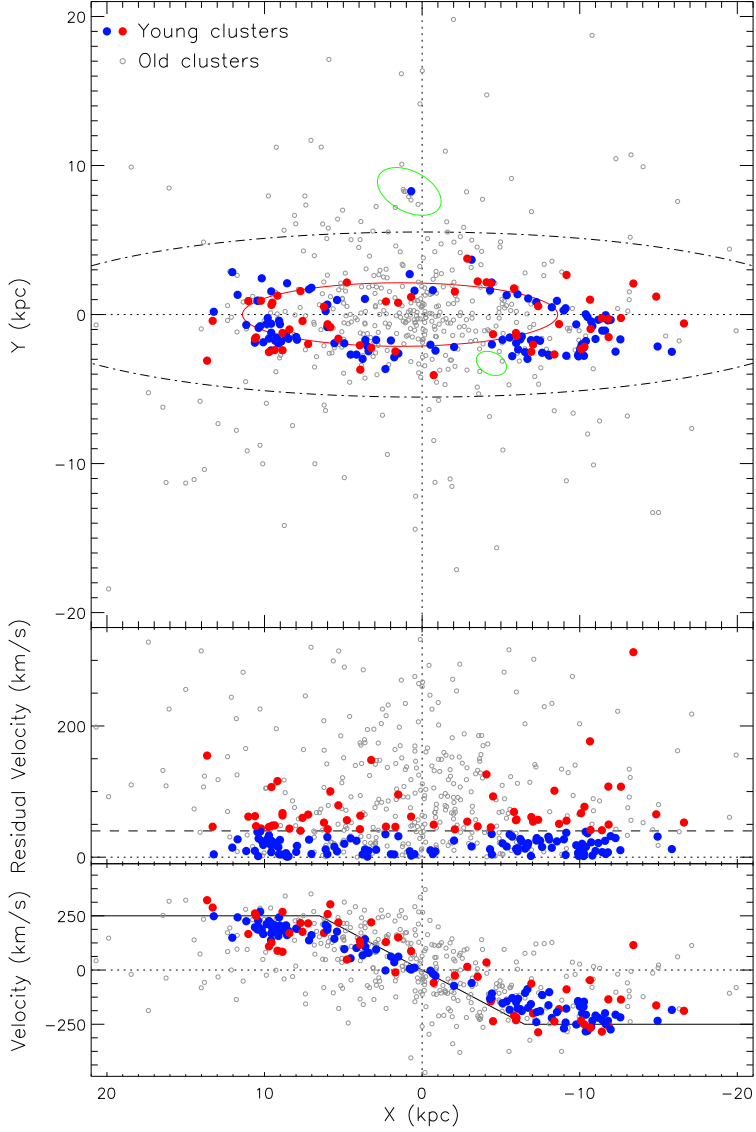}
\caption{
Spatial distribution (\textit{top panel}) and kinematical properties (\textit{middle and bottom panels}) of star clusters 
with available radial velocity. 
Colored filled circles are young clusters and small open circles are old clusters. 
(\textit{middle panel}) We present residual velocities defined as absolute values of the difference between calculated velocities from cold-disk model and 
observed velocities. 
Two subgroups are separated by residual velocity of 40~\kms (dashed horizontal line). 
(\textit{bottom panel}) Observed velocities corrected for the M31 system velocity against the projected distance along the major axis. 
The solid line is an adopted rotation curve that is flat with V$_{cirular}$ = 250~\kms for $|R|$ $>$ 6.5~kpc and then falls linearly to zero at $X$ = 0.
\label{fig16}}
\end{figure}
\clearpage

\begin{figure}
\includegraphics[width=120mm]{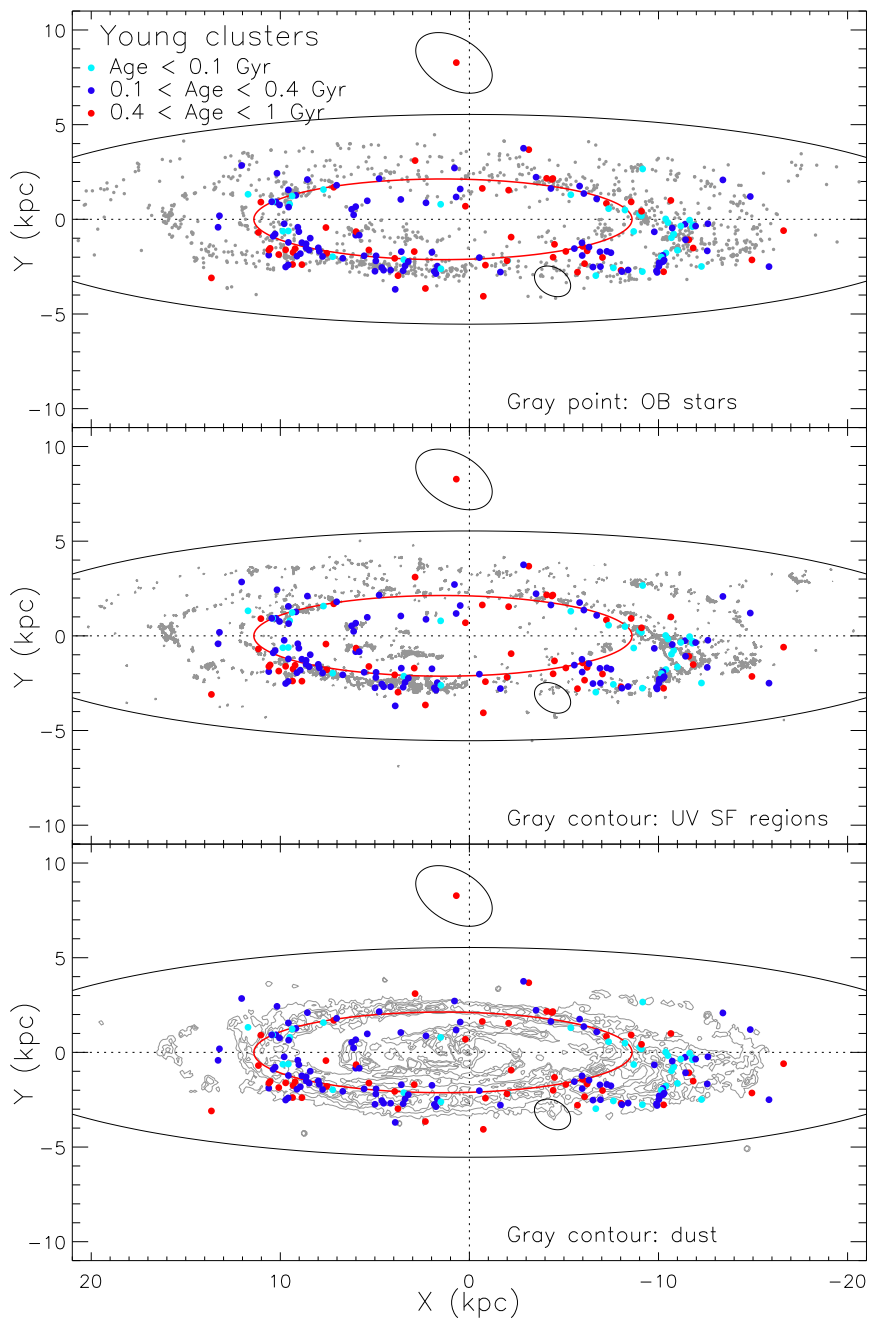}
\caption{
Spatial distribution of young clusters compared with OB stars (\textit{top panel}), 
$GALEX$ UV SF regions (\textit{middle panel}), and $Spitzer$ IRAC 8.0 $\micron$ IR contours (\textit{bottom panel}, courtesy P. Barmby). 
Filled circles with different colors are young clusters in different age ranges; 
age $<$ 100~Myr (cyan), 100~Myr $<$ age $<$ 400~Myr (blue), and 400~Myr $<$ age $<$ 1~Gyr (red).
\label{fig17}}
\end{figure}
\clearpage
 
\begin{figure}
\includegraphics[width=140mm]{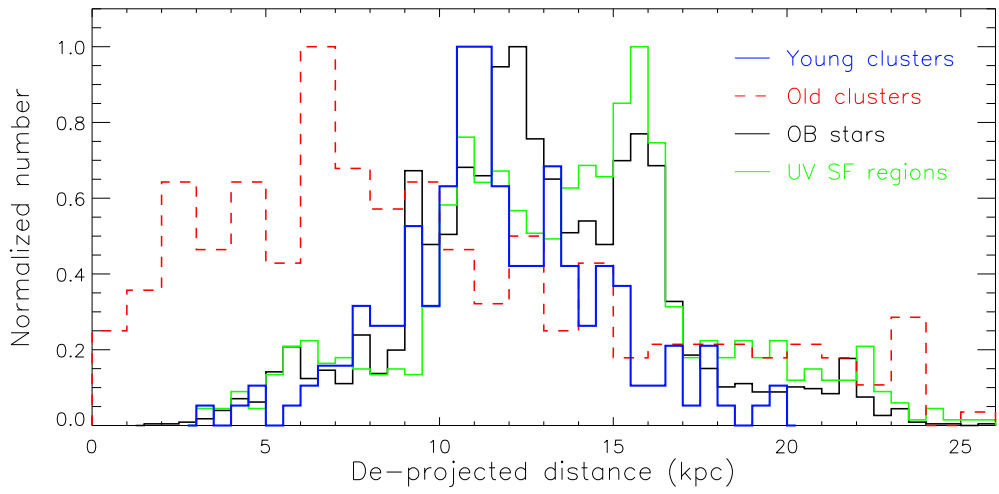}
\caption{
Number histogram of young clusters, old clusters, OB stars, and UV SF regions against de-projected distance from the center of M31. 
\label{fig18}}
\end{figure}
\clearpage

\begin{figure}
\includegraphics[width=97mm]{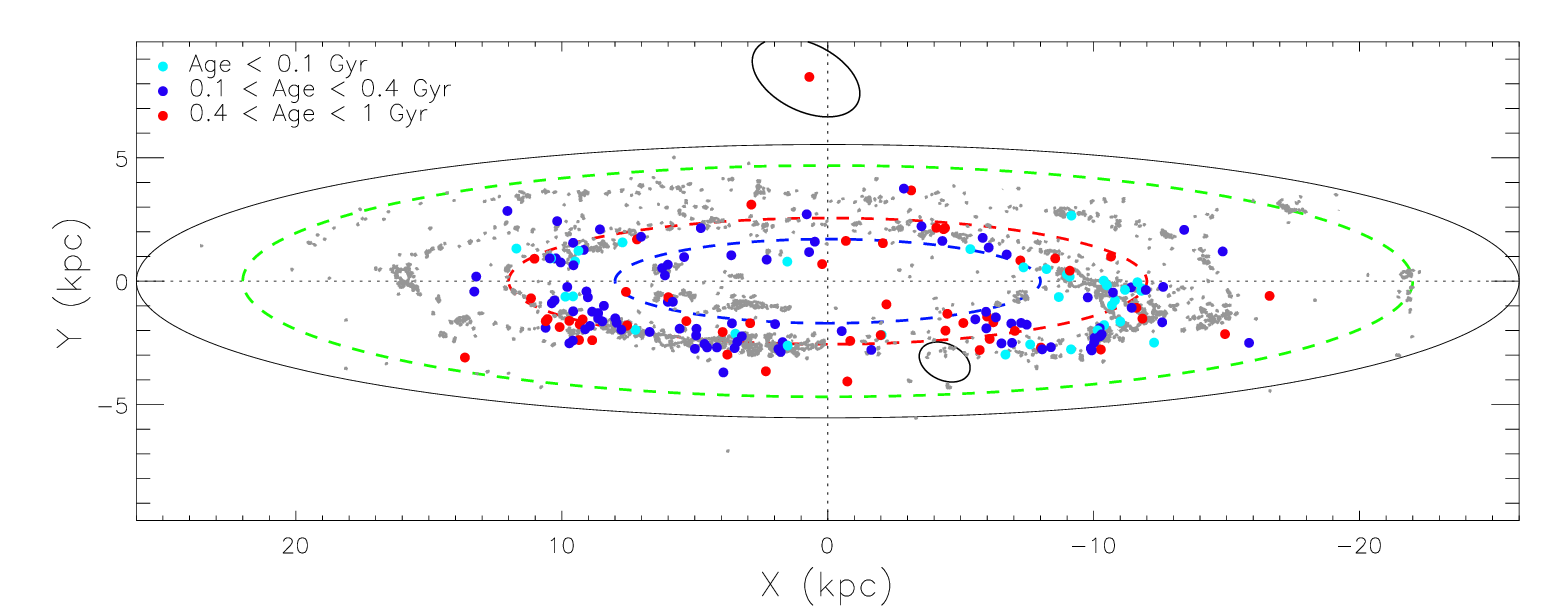}\\
\includegraphics[width=97mm]{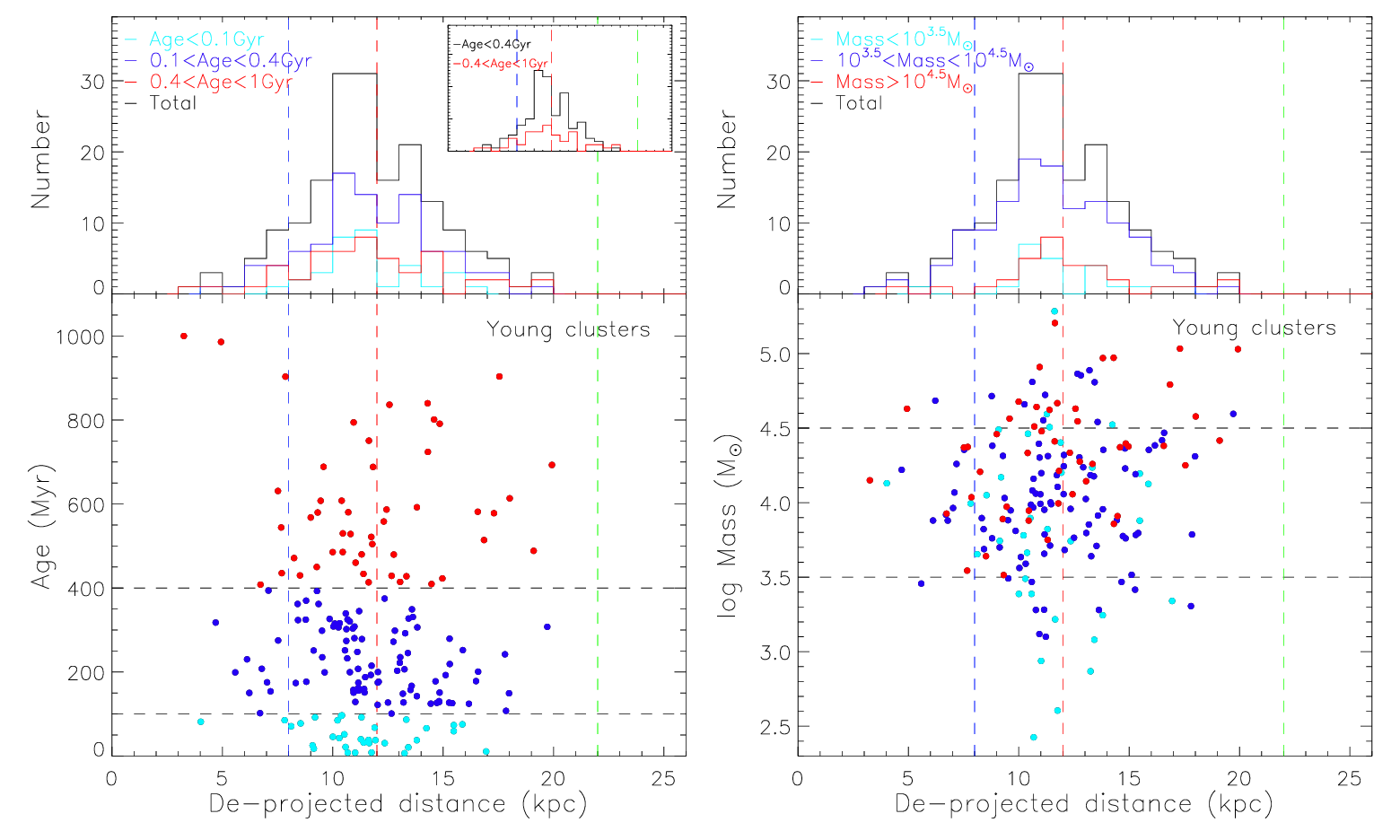}\\
\includegraphics[width=97mm]{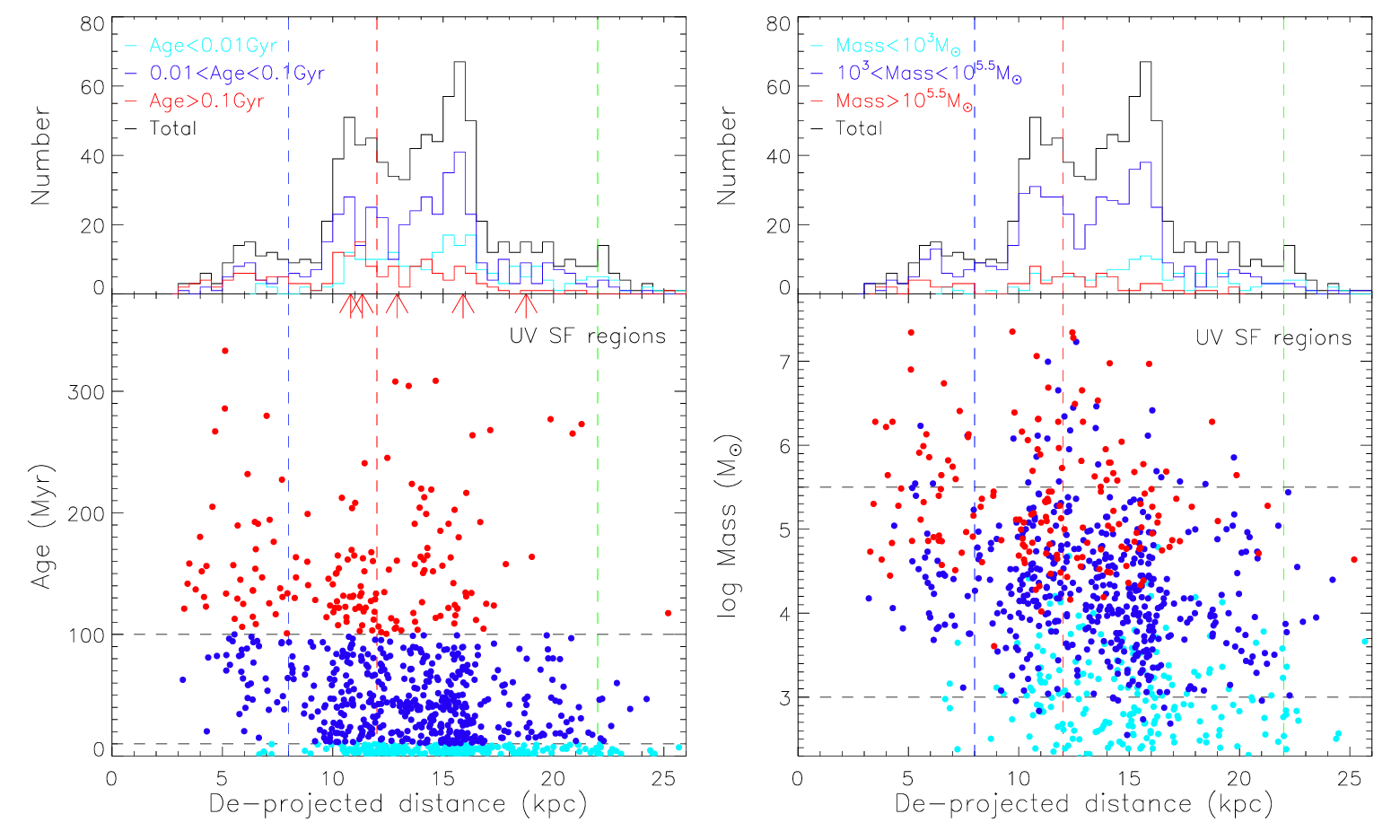}
\caption{
(\textit{Top panel}) The spatial distribution of young clusters (red filled circles) and $GALEX$ UV SF regions (yellow contours). 
(\textit{Middle panels}) Distribution of ages ($left$) and masses ($right$) of young clusters against de-projected 
distance from the center of M31. 
In the inset of the left panel, two subsamples with different age range (age $<$ 0.4~Gyr and 0.4~Gyr $<$ age $<$ 1~Gyr) 
are presented. 
(\textit{Bottom panels}) Distribution of ages ($left$) and masses ($right$) of UV SF regions against de-projected distance 
from the center of M31. 
The dashed vertical lines correspond to the ellipses drawn on the top panel at 8, 12, and 22~kpc, respectively. 
Dots with different colors indicate star clusters in different age ranges (red: $>$ 400~Myr, blue: 100 - 400~Myr, 
cyan: $<$ 100~Myr) and SF regions (red: $>$ 100~Myr, blue: 10 - 100~Myr, cyan: $<$ 10~Myr).
\label{fig19}}
\end{figure}
\clearpage
 
\begin{figure}
\includegraphics[width=70mm]{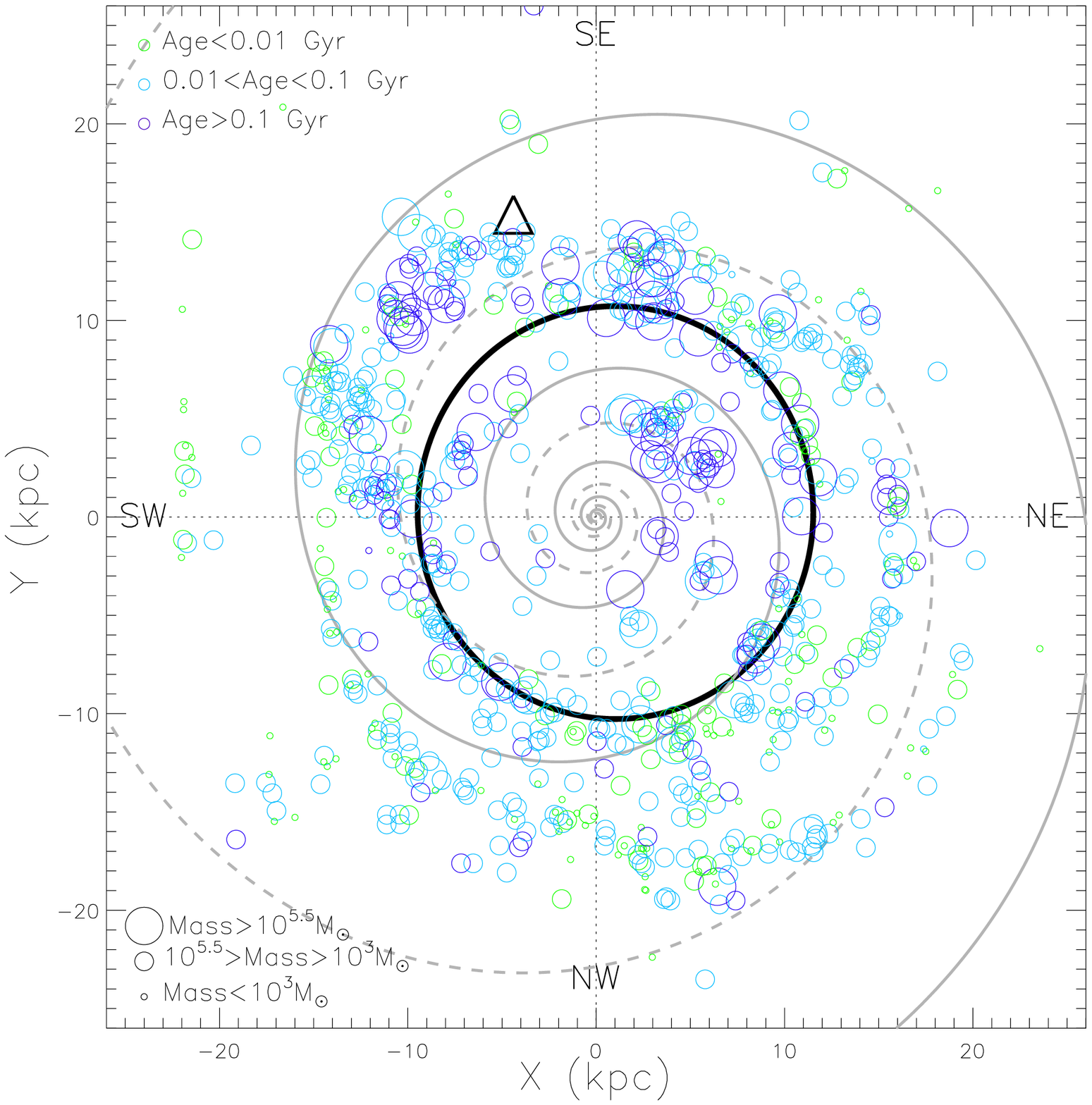}
\includegraphics[width=70mm]{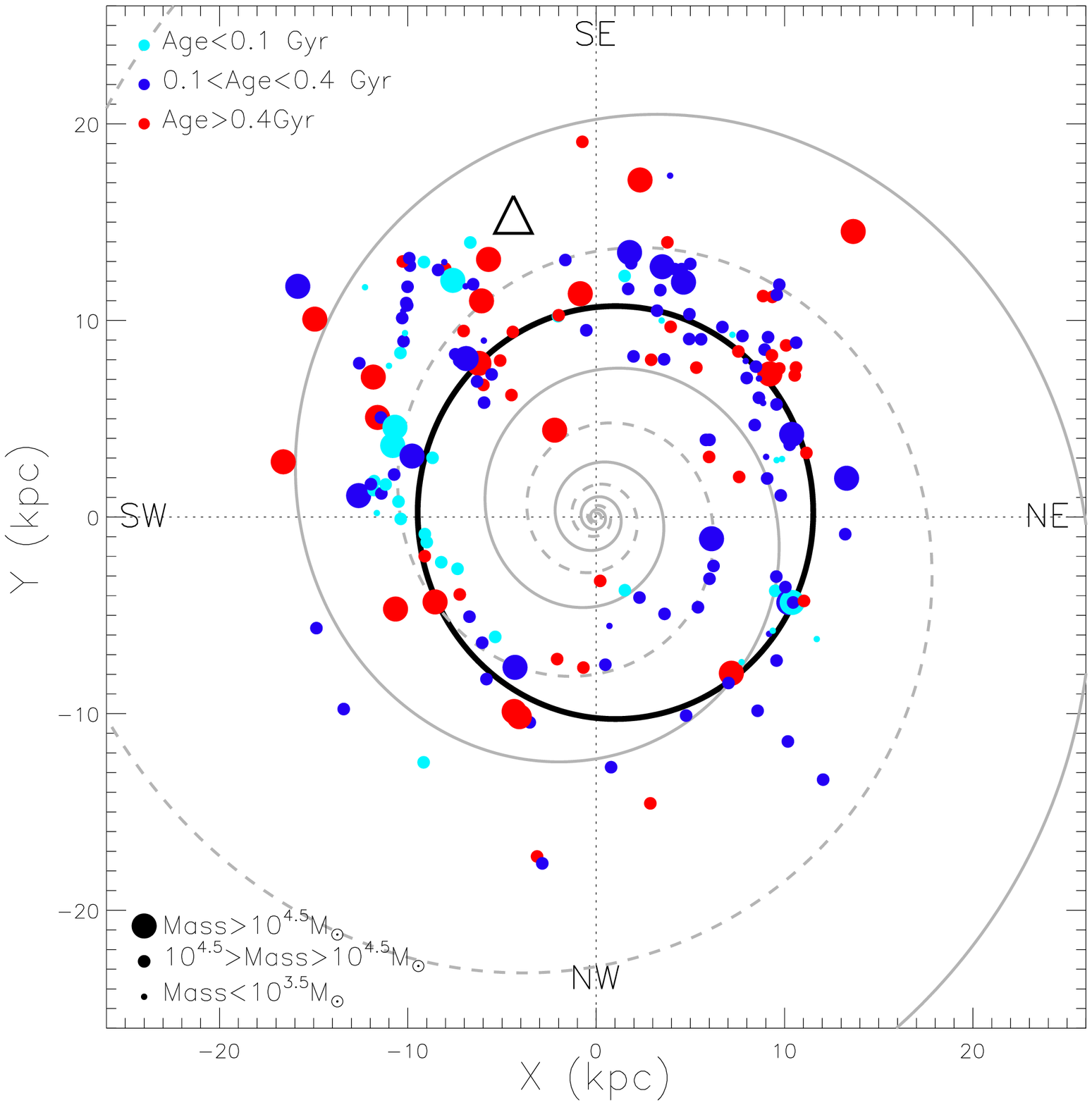}\\
\includegraphics[width=90mm]{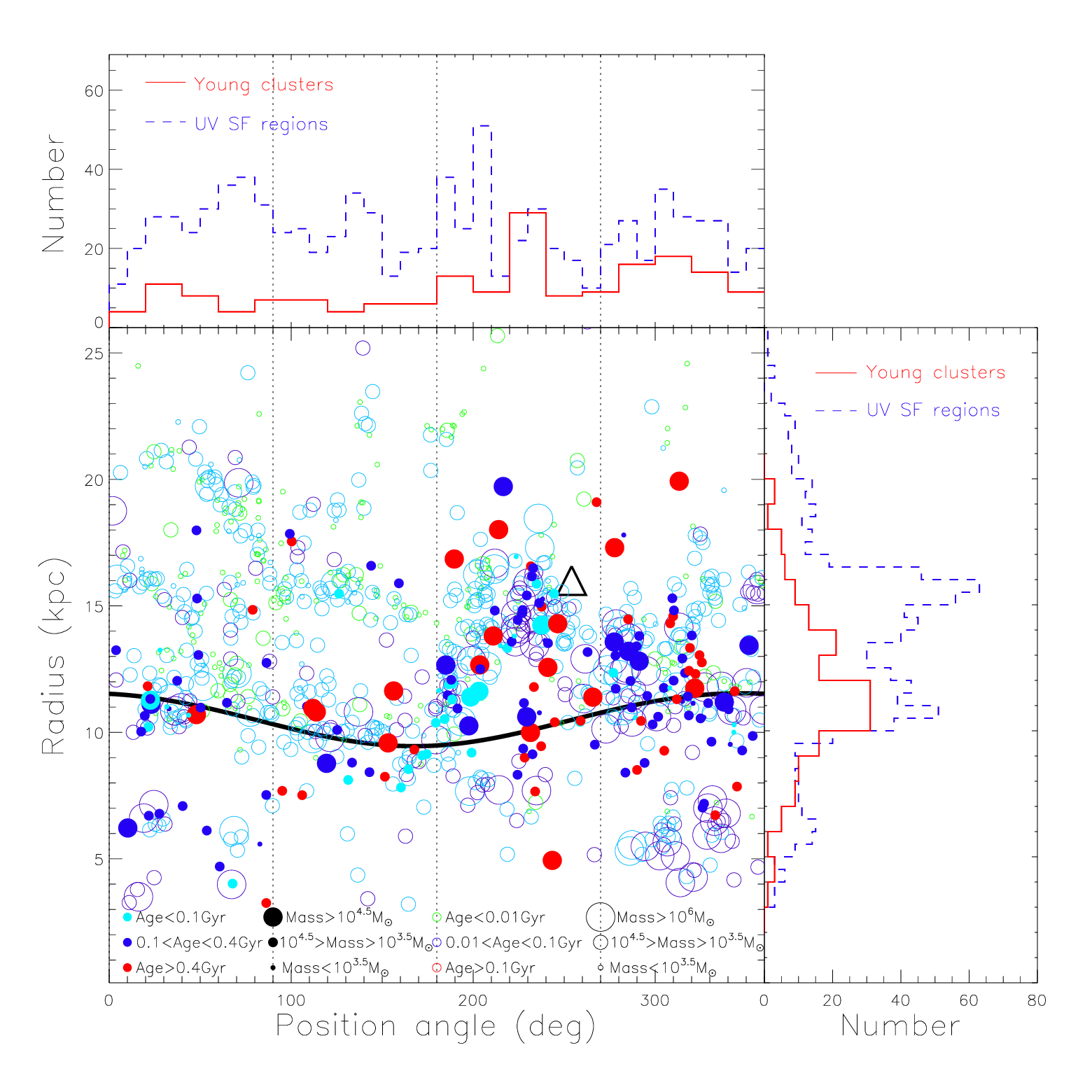}
\caption{
(\textit{Upper panels}) De-projected spatial distribution of UV SF regions (left) and young clusters (right). 
Different colors and sizes indicate objects in different age and mass ranges. 
The black large circle is the 10~kpc star formation ring and two gray spirals are simple logarithmic spiral arms adopted 
from \cite{gor06}. 
(\textit{Lower panel}) Spatial distribution of young clusters (filled circles) and SF regions (open circles) in polar coordinates. 
The black solid curve  is the 10~kpc circular star formation ring. 
The triangle in each panel marks the location of M32. 
\label{fig20}}
\end{figure}
\clearpage

\begin{figure}
\includegraphics[width=140mm]{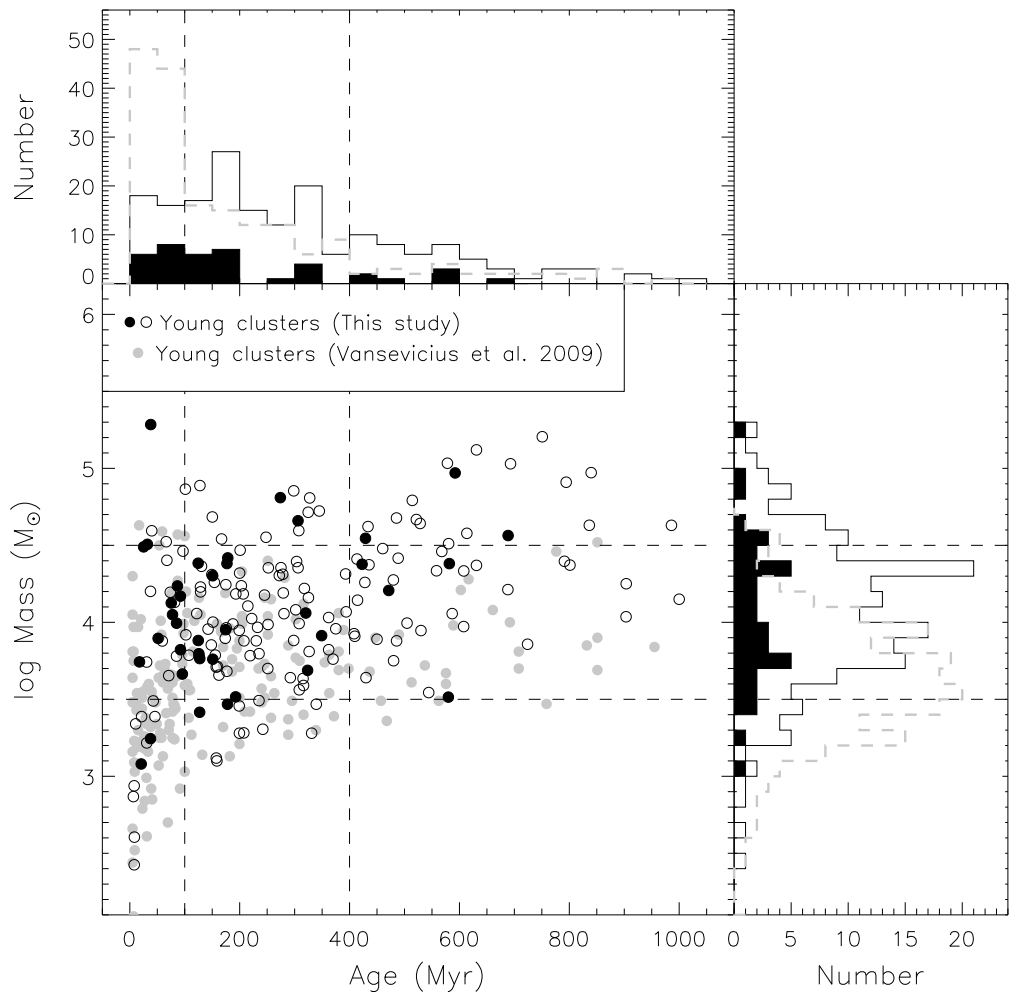}
\caption{
Age and mass distribution of our young cluster sample (black filled and open circles), and of compact star 
clusters (gray filled circles) from \cite{van09}. 
Black filled circles are objects in common between our sample and that of \cite{van09}. 
Age and mass histograms for different samples are also presented; all our young clusters (solid histogram), 
our young clusters in the field of \cite{van09} (filled histogram), and compact star clusters from \cite{van09} (gray dashed histogram). 
\label{fig21}}
\end{figure}
\clearpage

\begin{figure}
\includegraphics[width=140mm]{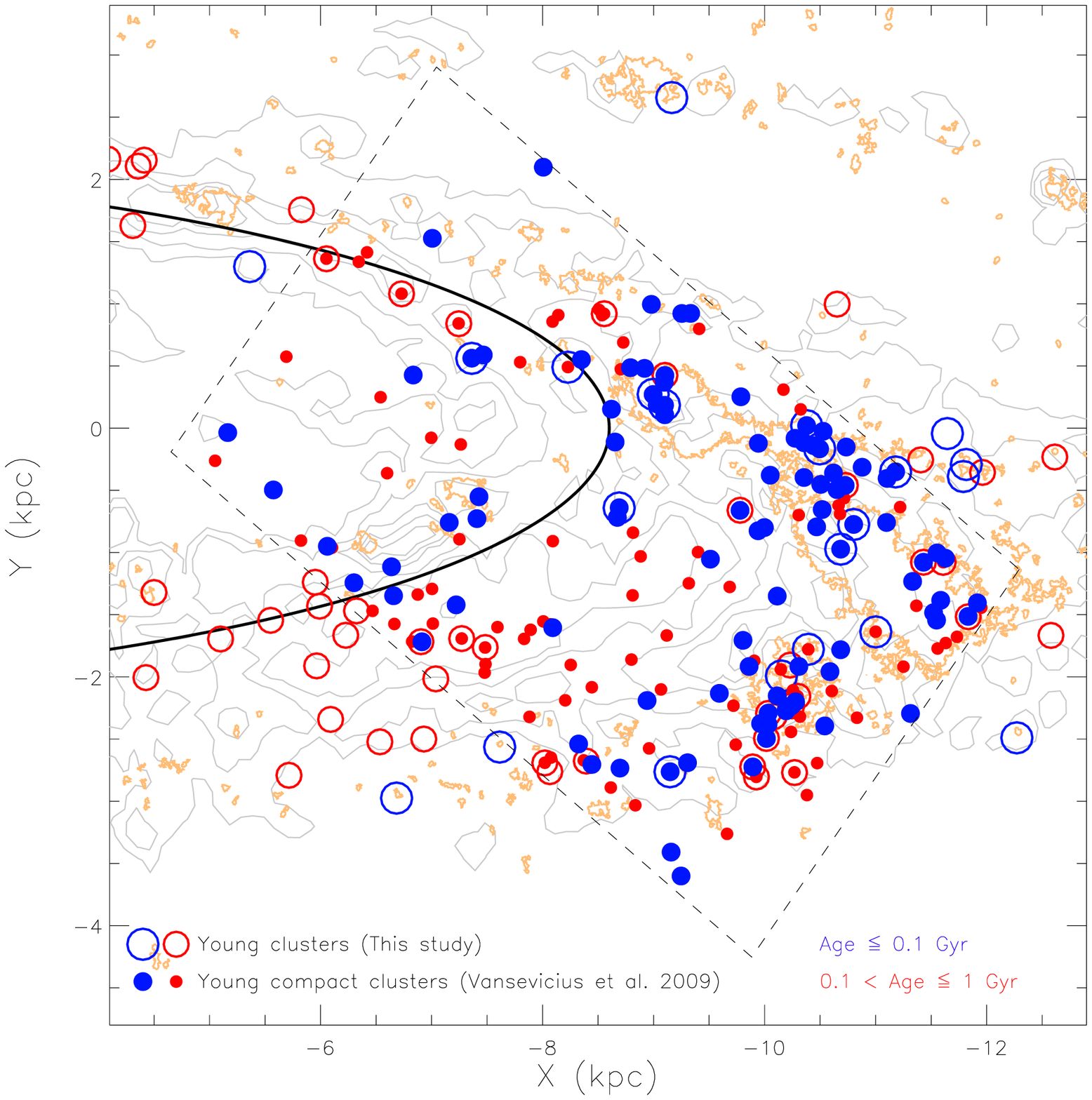}
\caption{
Spatial distribution of young clusters located in the southwestern part of the M31 disk. 
The large rectangle indicates the survey region of SUBARU Suprime-Cam by \cite{van09}. 
Filled circles are young ($\le$ 1~Gyr) compact star clusters from \cite{van09} and open circles are our young clusters. 
Orange contours are UV SF regions from \cite{kan09} and gray contours are for the dust distribution from the $Spitzer$ IRAC 8.0 $\micron$ image from \cite{bar06}. 
The large ellipse is the 10~kpc ring of the M31 disk.
\label{fig22}}
\end{figure}
\clearpage


\begin{flushleft}
\begin{deluxetable}{ccccccccccccccccccccccccccccccc}
\rotate
\setlength{\tabcolsep}{0.01in}
\tabletypesize{\tiny}
\tablecolumns{31}
\tablewidth{0pc}
\tablecaption{Merged catalog of star clusters in M31.}
\tablehead{
\colhead{Name} & \colhead{RA} & \colhead{DEC} & \colhead{FUV} & \colhead{$\sigma_{FUV}$} & \colhead{NUV} & \colhead{$\sigma_{NUV}$} & \colhead{$U$} & \colhead{$B$} & \colhead{$V$} & \colhead{$R$} & \colhead{$I$} & \colhead{$J$} & \colhead{$H$} & \colhead{$K_{RBC}$} & \colhead{$u$} & \colhead{$g$} & \colhead{$r$} & \colhead{$i$} & \colhead{$z$} & \colhead{$K_{P10}$} & \colhead{$E(B-V)$} & \colhead{$\sigma_{E(B-V)}$} & \colhead{[Fe/H]\tablenotemark{a}} & \colhead{$\sigma_{[Fe/H]}$} & \colhead{V$_{r}$} & \colhead{$\sigma_{V_{r}}$} & \colhead{$f_{RBC}$} &  \colhead{$f_{P10}$} &  \colhead{$f_{C11}$} &  \colhead{$f_{EBV}$}}
\startdata
      B001 &   00:39:51.01 &  40:58:10.6 &\nodata&\nodata&\nodata&\nodata&  18.82 &  18.33 &  17.06 &  16.47 &  15.41 &  14.68 &  13.73 &  13.86 &  19.38 &  17.58 &  16.61 &  16.07 &  15.69 &  13.72 &   0.25 &   0.02 &  -0.42 &   0.32 &  -191.1 &   14.3 &   1 &   1 &   3 &   1 \\
      B002 &   00:40:02.58 &  41:11:53.5 &\nodata&\nodata&  21.28 &   0.04 &  18.14 &  18.18 &  17.55 &  17.12 &  16.58 &  14.87 &  14.77 &\nodata&  19.15 &  17.86 &  17.34 &  17.06 &  16.90 &  15.47 &   0.01 &   0.04 &  99.99 &  99.99 &  -338.2 &   14.5 &   1 &   1 &   3 &   1 \\
      B003 &   00:40:09.41 &  41:11:05.7 &  23.06 &   0.20 &  21.55 &   0.04 &  18.40 &  18.35 &  17.57 &  17.07 &  16.41 &  15.96 &  15.16 &  15.54 &  19.43 &  17.94 &  17.36 &  16.99 &  16.82 &  15.09 &   0.16 &   0.04 &  -0.99 &   0.48 &  -364.0 &   15.3 &   1 &   1 &   3 &   1 \\
      B004 &   00:40:17.92 &  41:22:40.3 &\nodata&\nodata&  22.25 &   0.07 &  18.29 &  17.87 &  16.95 &  16.36 &  15.73 &  14.96 &  14.24 &  14.10 &  19.06 &  17.40 &  16.64 &  16.27 &  16.05 &  14.17 &   0.13 &   0.03 &  -1.00 &   0.41 &  -369.4 &   13.4 &   1 &   1 &   3 &   1 \\
      B005 &   00:40:20.32 &  40:43:58.3 &\nodata&\nodata&  21.03 &   0.07 &  16.85 &  16.64 &  15.69 &  15.02 &  14.40 &  13.39 &  12.68 &  12.53 &  17.87 &  16.12 &  15.32 &  14.90 &  14.62 &  12.56 &   0.22 &   0.05 &  -0.82 &   0.38 &  -278.3 &   12.8 &   1 &   1 &   3 &   1 \\
      B006 &   00:40:26.49 &  41:27:26.7 &\nodata&\nodata&  21.41 &   0.04 &  16.94 &  16.49 &  15.50 &  14.97 &  14.33 &  13.47 &  12.74 &  12.63 &  17.68 &  15.92 &  15.16 &  14.78 &  14.52 &  12.55 &   0.11 &   0.03 &  -0.59 &   0.41 &  -234.7 &    5.8 &   1 &   1 &   3 &   1 \\
      B008 &   00:40:30.29 &  41:16:08.7 &\nodata&\nodata&  22.59 &   0.12 &  18.16 &  17.66 &  16.56 &  16.21 &  15.51 &  14.66 &  14.05 &  13.89 &  19.03 &  17.23 &  16.47 &  16.07 &  15.85 &  14.02 &   0.17 &   0.09 &  -0.47 &   0.35 &  -318.5 &   12.6 &   1 &   1 &   3 &   2 \\
      B009 &   00:40:30.70 &  41:36:55.6 &  23.09 &   0.15 &  20.96 &   0.03 &  17.54 &  17.63 &  16.92 &  16.42 &  15.87 &  15.27 &  14.47 &  14.42 &  18.55 &  17.24 &  16.63 &  16.33 &  16.19 &  14.67 &   0.09 &   0.04 &  -1.55 &   0.23 &  -325.5 &   52.9 &   1 &   1 &   3 &   1 \\
      B010 &   00:40:31.57 &  41:14:22.5 &  21.93 &   0.08 &  20.87 &   0.03 &  17.65 &  17.50 &  16.66 &  16.12 &  15.48 &  14.83 &  14.28 &  13.98 &  18.45 &  17.03 &  16.38 &  16.02 &  15.80 &  14.16 &   0.20 &   0.03 &  -1.64 &   0.68 &  -162.7 &   14.7 &   1 &   1 &   3 &   1 \\
      B011 &   00:40:31.88 &  41:39:16.9 &  21.84 &   0.06 &  20.58 &   0.02 &  17.59 &  17.39 &  16.58 &  16.10 &  15.56 &  14.85 &  14.23 &  14.08 &  18.42 &  17.06 &  16.43 &  16.14 &  15.97 &  14.17 &   0.09 &   0.04 &  -1.71 &   0.24 &  -207.9 &   53.7 &   1 &   1 &   3 &   1 \\
      B012 &   00:40:32.47 &  41:21:44.2 &  20.10 &   0.02 &  19.02 &   0.01 &  15.99 &  15.86 &  15.09 &  14.62 &  14.03 &  13.36 &  12.78 &  12.74 &  16.76 &  15.43 &  14.83 &  14.52 &  14.35 &  12.73 &   0.11 &   0.02 &  -1.91 &   0.21 &  -359.4 &   11.3 &   1 &   1 &   3 &   1 \\
      B013 &   00:40:38.44 &  41:25:23.7 &\nodata&\nodata&  23.21 &   0.25 &  18.56 &  18.06 &  17.19 &  16.60 &  15.96 &  15.22 &  14.46 &  14.22 &  19.19 &  17.63 &  16.88 &  16.47 &  16.19 &  14.44 &   0.13 &   0.06 &  -0.74 &   0.51 &  -410.2 &   13.2 &   1 &   1 &   3 &   1 \\
      B015 &   00:40:45.03 &  40:59:56.1 &\nodata&\nodata&\nodata&\nodata&\nodata&  19.20 &  17.79 &  16.93 &  15.90 &  14.61 &  13.75 &  13.44 &  20.74 &  18.61 &  17.43 &  16.70 &  16.26 &  13.67 &   0.61 &   0.03 &   0.37 &   0.02 &  -460.0 &   14.0 &   1 &   1 &   3 &   1 \\
      B016 &   00:40:45.17 &  41:22:09.9 &\nodata&\nodata&  23.46 &   0.24 &  18.86 &  18.58 &  17.58 &  16.85 &  16.15 &  15.15 &  14.17 &  14.08 &  19.98 &  17.99 &  17.18 &  16.74 &  16.42 &  14.44 &   0.35 &   0.04 &  -0.53 &   0.34 &  -397.2 &   13.2 &   1 &   1 &   3 &   1 \\
      B017 &   00:40:48.72 &  41:12:07.2 &\nodata&\nodata&  21.57 &   0.13 &  17.55 &  17.04 &  15.95 &  15.23 &  14.51 &  13.47 &  12.69 &  12.60 &  18.27 &  16.48 &  15.51 &  14.97 &  14.59 &  12.53 &   0.32 &   0.03 &  -0.82 &   0.24 &  -522.2 &    9.8 &   1 &   1 &   3 &   1 \\
 \nodata  &  \nodata  &  \nodata  &  \nodata  &  \nodata  &  \nodata  &  \nodata  &  \nodata  &  \nodata  &  \nodata  &  \nodata  &  \nodata  &  \nodata  &  \nodata  &  \nodata  &  \nodata  &  \nodata  &  \nodata  &  \nodata  &  \nodata  &  \nodata  &  \nodata  &  \nodata  &  \nodata  &  \nodata  &  \nodata  &  \nodata  &  \nodata  &  \nodata  &  \nodata  &  \nodata  \\
\enddata
\tablecomments{This table is available in its entirety in the electronic edition of the online journal. 
A portion is shown here for guidance regarding its format and content.}
\tablenotetext{a}{Values in parentheses are either from \cite{bea04} or \cite{per10}, or the mean of these values (for B222, B321, and B327).}
\end{deluxetable}
\end{flushleft}

\begin{deluxetable}{ccccccccc}
\tablecolumns{10}   
\tabletypesize{\small}
\tablewidth{0pc} 
\tablecaption{Catalog of 182 young clusters.}
\tablehead{  
\colhead{Name} & \colhead{RA} & \colhead{DEC} & \colhead{Age} & \colhead{Log Mass} & \colhead{$E(B-V)$} & \colhead{$f_{EBV}$\tablenotemark{a}} & \colhead{Z\tablenotemark{b}} & \colhead{Flag\tablenotemark{c}}\\
\colhead{} & \colhead{(hh:mm:ss)} & \colhead{(dd:mm:ss)} & \colhead{(Myr)} & \colhead{(M$_{\sun}$)} & \colhead{(mag)} & \colhead{} & \colhead{} & \colhead{}}
\startdata
      B040 &   0:41:38.86 &  40:40:54.4 &   422 &  4.38 &   0.00 &   2 &  0.004 &   1 \\
      B043 &   0:41:42.31 &  40:42:39.8 &    66 &  4.52 &   0.24 &   1 &   0.05 &   1 \\
      B049 &   0:41:45.58 &  40:49:55.0 &   485 &  4.68 &   0.21 &   1 &   0.02 &   1 \\
      B066 &   0:42:03.09 &  40:44:47.1 &    73 &  4.20 &   0.14 &   1 &   0.05 &   1 \\
      B069 &   0:42:05.52 &  41:26:09.2 &   272 &  4.30 &   0.19 &   1 &   0.05 &   1 \\
      B081 &   0:42:13.59 &  40:48:39.0 &   839 &  4.97 &   0.17 &   2 &  0.004 &   1 \\
      B091 &   0:42:21.71 &  41:22:05.3 &   275 &  4.35 &   0.11 &   2 &   0.05 &   1 \\
      B133 &   0:42:51.60 &  41:23:29.7 &    81 &  4.13 &   0.01 &   2 &   0.05 &   2 \\
      B192 &   0:43:44.52 &  41:37:27.0 &   393 &  4.07 &   0.00 &   2 &   0.02 &   1 \\
      B195 &   0:43:48.55 &  41:02:27.9 &   488 &  4.42 &   0.36 &   2 & 0.0004 &   1 \\
      \nodata &  \nodata &  \nodata &   \nodata &    \nodata &  \nodata &  \nodata &  \nodata &  \nodata \\
\enddata
\tablecomments{This table is available in its entirety in the electronic edition of the online journal.
A portion is shown here for guidance regarding its format and content.}
\tablenotetext{a}{Flag of $E(B-V)$: (1) reddening value from our compiled catalog ($indivEBV$), (2) reddening value from our SED fitting ($freeEBV$).} 
\tablenotetext{b}{Metallicity adopted in SED fitting (see Section 3.1).}
\tablenotetext{c}{Young clusters flag: (1) young clusters in both this work and \cite{cal09}, 
(2) only in this work, 
(3) only in \cite{cal09}.}
\end{deluxetable}
\clearpage

\end{document}